\documentclass[10pt,journal,twocolumn]{IEEEtran}

\IEEEoverridecommandlockouts
\usepackage{cite}
\usepackage{amsmath,amssymb,amsfonts,booktabs}
\usepackage{algorithmic}
\usepackage{graphicx}
\usepackage{textcomp}
\usepackage{xcolor}
\usepackage{enumerate}
\usepackage{epstopdf}
\usepackage{subcaption}
\usepackage{amsthm}
\newtheorem{theorem}{Theorem}
\newtheorem{lemma}{Lemma}
\newtheorem{proposition}{Proposition}
\newtheorem{definition}{Definition}
\newtheorem{rmk}{Remark}

\usepackage{multirow} 

\definecolor{mygray}{RGB}{240,240,240}
\usepackage{colortbl}

\usepackage{algorithm}
\usepackage{algorithmic}
\allowdisplaybreaks[4]

\begin{document}

\title{\huge Joint Flexible Beamforming, Resource Allocation, and Task Buffering for Movable-Antenna Enabled MEC Systems}

\author{
Hao Wang, Chunjie Wang, Lingtong Ma, Wenchao Liu, Kan Yu, Jinke Ren, Yanyan Shen, and Xuhui Zhang

\thanks{
Hao Wang and Lingtong Ma are with the Department of Networks, China Mobile Communications Group Co., Ltd., Beijing 100033, China (e-mail: wanghaowl@chinamobile.com; malingtong@chinamobile.com).
}

\thanks{
Chunjie Wang is with Shenzhen Institutes of Advanced Technology, Chinese Academy of Sciences, Guangdong 518055, China, and also with the University of Chinese Academy of Sciences, Beijing 100049, China (e-mail: cj.wang@siat.ac.cn).
}

\thanks{
Wenchao Liu is with the School of Automation and Intelligent Manufacturing, Southern University of Science and Technology, Shenzhen 518055, China (e-mail: wc.liu@foxmail.com).
}

\thanks{
Kan Yu is with
the Key Laboratory of Universal Wireless Communications, Ministry of Education,
Beijing University of Posts and Telecommunications, Beijing, 100876, China (e-mail: kanyu1108@126.com).
}

\thanks{
Jinke Ren and Xuhui Zhang are with the Shenzhen Future Network of Intelligence Institute, the School of Science and Engineering, and the Guangdong Provincial Key Laboratory of Future Networks of Intelligence, The Chinese University of Hong Kong, Shenzhen, Guangdong 518172, China (e-mail: jinkeren@cuhk.edu.cn; xu.hui.zhang@foxmail.com).
}

\thanks{
Yanyan Shen is with Shenzhen Institutes of Advanced Technology, Chinese Academy of Sciences, Guangdong 518055, China (e-mail: yy.shen@siat.ac.cn).
}

}

\maketitle

\begin{abstract}
The proliferation of computation-intensive applications, such as generative artificial intelligence and low-altitude economy services, demands mobile edge computing (MEC) systems that are both energy-efficient and adaptable to highly dynamic workloads. However, conventional MEC architectures rely on fixed-position antennas with limited spatial flexibility and rigid single-slot task execution models, which fail to accommodate bursty traffic and time-varying channels. To address these challenges, this paper proposes a novel MEC framework empowered by movable antennas (MAs) and task buffering. Unlike traditional approaches, our system enables continuous antenna repositioning to maximize array gains and decouples task generation from immediate execution via buffer queues, allowing for flexible multi-slot scheduling. We formulate a stochastic optimization problem to minimize long-term energy consumption by jointly optimizing receive beamforming, MA positions, and computing resources. Leveraging Lyapunov optimization, we transform the long-term problem into tractable per-slot subproblems and develop an efficient alternating optimization algorithm. Simulation results demonstrate that the proposed scheme significantly outperforms the benchmark schemes in terms of energy consumption, while maintaining stable task queues under unpredictable traffic conditions.
\end{abstract}


\section{Introduction}
Mobile edge computing (MEC) has rapidly evolved from a simple latency-aware computation mechanism into the cornerstone of next-generation intelligent network infrastructures \cite{9252924, 10422876}.
In recent years, its significance has been amplified by the surge in computationally demanding applications, such as service-oriented generative artificial intelligence (AI), blockchain-enabled intelligent computing \cite{11106440, liu2025computing, 10628028} and the burgeoning low-altitude economy (LAE) applications for Internet of things (IoTs) and uncrewed aerial vehicles (UAVs) \cite{wu2025low, wang2023incentive}.
Specifically, generative AI services require massive, real-time inference capabilities that traditional cloud-only architectures cannot support due to backhaul constraints, while emerging services, such as autonomous vehicle trajectory design and urban aerial mobility detection, demand ultra-reliable, low-latency communication for safe navigation and data processing \cite{10734319}.
MEC addresses these challenges by decentralizing computing resources, bringing them closer to end-users to facilitate efficient task offloading, dynamic resource allocation, and robust task data management \cite{9214878, 10012694}.

Despite these architectural advances, the physical layer of MEC systems remains a critical bottleneck. Traditional single fixed position antenna (FPA) systems are fundamentally limited by poor spatial diversity and are susceptible to channel fading, leading to unstable transmission rates that undermine the quality of services (QoEs) \cite{10741192, 11154927}.
In contrast, conventional multi-FPA systems offer significant improvements through spatial multiplexing and beamforming gains, which enhance spectral efficiency and link reliability \cite{8766143}. However, the FPA arrays lack the agility to dynamically reconfigure their radiation patterns or physical aperture in response to highly mobile users and rapidly time-varying channel conditions \cite{11007274, liu2025movable}.
This structural rigidity restricts the system's ability to fully exploit spatial degrees of freedom (DoFs) in dynamic environments, ultimately capping the throughput and reliability required for computation-intensive MEC services \cite{xiao2025fundamental, 11418663}.

Recently, research focus is shifting towards movable antenna (MA) technology to overcome the spatial constraints of fixed architectures \cite{11226954, 11543384, 11177504}.
Unlike FPA arrays, MAs are capable of dynamically repositioning antenna elements within a designated region, enabling precise adaptation to channel variations \cite{10693833}.
The mobility allows the system to continuously seek optimized signal strengths by aligning antenna positions with favorable channel phases, thereby maximizing array gains and spatial diversity without increasing the number of antennas.
Furthermore, MAs provide enhanced interference suppression capabilities by flexibly adjusting element positions to nullify interfering signals \cite{11346858, 11374005}.
By bridging the gap between hardware flexibility and channel dynamics to improve spectral efficiency and link reliability, MAs are a suitable enabler for the stringent requirements of dynamic MEC environments \cite{11517458}.

Although MAs provide the precise and adaptive transmission, existing task offloading strategies often operate under a restrictive assumption, mandating that all tasks generated within a time slot be completed immediately within the slot \cite{9968198}.
However, this single-slot constraint is too rigid for practical systems \cite{10896833}. Workloads from bursty generative AI or LAE services are not only highly dynamic and unpredictable but also require the flexibility to be processed continuously across multiple time slots to ensure successful completion with efficient resource utilization and decreased task drop rates.
To address this, a more flexible paradigm is required that decouples task generation from immediate execution through task buffering.
Specifically, by caching tasks in buffer queues, the system gains the freedom to schedule computations over the rest of the time interval, shifting workloads from congested periods to idle slots for balanced energy-latency trade-offs \cite{11197291}.
To rigorously manage these buffering dynamics without requiring prior knowledge of stochastic task arrivals or environmental information, Lyapunov optimization emerges as a powerful framework \cite{9968198}.
By constructing virtual buffer queues, it facilitates online decision-making that stabilizes the system while optimizing long-term performance, transforming the MEC system into a buffer queue-aware system capable of efficiently handling unpredictable demands.

Motivated by the urgent requirements to support latency-sensitive generative AI and LAE services, this paper investigates an MA-enabled MEC system where both mobile users (MUs) and the base station (BS) buffer computation tasks in task buffer queues for flexible scheduling.
Unlike traditional FPA-based MEC systems that have a one-slot deadline, we utilize MAs to dynamically adjust antenna positions to optimize task offloading rates of all MUs. Lyapunov optimization is utilized to decouple the long-term energy minimization problem into per-slot subproblems.
Then, we propose an alternating optimization (AO) algorithm to solve the subproblems.
Our main contributions are presented as follows:
\begin{itemize}
    \item We consider an MA-enabled MEC system providing task computing services for multiple MUs.
    Task buffers are considered at both the BS and MUs for efficient and sustainable task computing.
    We formulate a stochastic optimization problem to jointly optimize the beamforming, the MA positions, and the computing resource, aiming at minimizing the long-term accumulated energy consumption of both the BS and MUs.
    \item We transform the long-term stochastic problem into deterministic per-slot optimization subproblems by the Lyapunov optimization. Then, an AO algorithm is designed to iteratively optimize the receive beamforming, the MA position, and the computing resource allocation.
    We rigorously analyze the computational complexity of the AO algorithm.
    \item Extensive simulations are conducted to validate the proposed scheme. Simulation results demonstrate that the MA-enabled scheme significantly outperforms the FPA-based scheme in terms of energy consumption and the sum computed data.
    The stability of the constructed task buffers and the long-term energy consumption are also confirmed under different weighted parameters.
\end{itemize}

\textit{Organizations:}
The rest of this paper is presented as follows.
Section II provides a comprehensive review of previous work.
In Section III, we establish the system model for the MA-enabled MEC system incorporating task buffering, and subsequently formulate the long-term energy consumption minimization problem.
Then, in Section IV, we first transform the original problem into a tractable per-slot form, and then propose an AO algorithm to solve it, accompanied by the analysis of computational complexity.
In Section V, extensive simulation results are presented to validate the proposed scheme, demonstrating its superiority in terms of energy consumption, task processing capability, and the stability of the task buffer queues.
Finally, Section VI concludes the paper.

\section{Related Works}

To provide a structured overview of the related literature, this section is organized along three research directions. First, we review the MA-enabled wireless systems, with emphasis on downlink transmission scenarios.
Second, we discuss the emerging research on the MA-enabled MEC systems, highlighting the gap in task offloading designs with flexible task scheduling.
Third, we survey the MEC systems with task-oriented storage mechanisms, encompassing both task caching and task buffering approaches.
The detailed review enables us to position our contributions at the intersection of these three directions, addressing the underexplored integration of the MA technology with multi-slot task buffering for MEC systems.

\subsection{MA-Enabled Systems for Downlink Transmissions}
MAs, also known as fluid antennas \cite{11328802}, have recently garnered significant attention in wireless communications as a promising solution to overcome the spatial constraints of the FPA arrays \cite{11281876, 11494046}.
A general channel estimation (CE) scheme was studied in \cite{10497534}, providing a compressed sensing based CE for MA systems.
In \cite{10354003}, an MA-enabled multiple-access system for uplink transmission was investigated to reduce the total transmit power of users.
The downloading throughput capacity of an MA-enhanced LAE system was optimized by designing the transmit beamforming, the antenna position, and the UAV trajectory \cite{10654366}.
To improve the physical layer security, the secrecy rate performance enhanced by MAs was optimized without the accurate location of eavesdroppers \cite{11156108}.
The energy efficiency of a general multi-antenna system assisted by MAs was optimized under different power constraint cases with a proposed low-complexity solution \cite{11261377}.
Moreover, the MA-enabled integrated sensing and communication (ISAC) for the vehicular network was investigated \cite{11363226}, and the sensing and communication performance were improved by optimizing communication resource allocation and antenna positions.
The recent advancements in communication performance improved by MAs have inspired us to explore the integration of MA technology into MEC systems by fully utilizing the spatial DoFs of MAs to realize flexible beamforming.

\subsection{MA-Enabled MEC Systems}
While MA technology has demonstrated impressive gains in signal enhancement and interference mitigation, existing research has mainly focused on downlink transmission scenarios, with limited attention paid to uplink transmission \cite{11240557}.
This gap is particularly critical for MEC architectures, which fundamentally rely on efficient uplink communication from users to offload computation tasks to edge servers.
Several pioneering works have started to bridge this gap.
For instance, the task completion latency was minimized in the MA-enabled MEC systems \cite{10528324, 11430678}.
Subsequently, the task completion efficiency was investigated in the ISAC-MEC system \cite{11017619},
the device-to-device aided MEC system \cite{11098592}, the satellite-assisted MEC system \cite{11301741}, and the anti-jamming MEC system \cite{11442647}, respectively.
Moreover, the wireless powered MEC system was studied \cite{10620306}, where the BS provided energy and computing services for users through antenna position optimization and resource allocation.
In \cite{10971386}, the discrete antenna position optimization was considered in the MA-enabled MEC system with the minimum antenna distance constraint.
However, existing MA-enabled MEC systems often assume that the MEC servers perform only immediate, simple task processing services, overlooking the urgent need for advanced data buffering and intelligent scheduling capabilities to handle increasingly complex computation-extensive tasks.

\subsection{{MEC Systems with Task-Oriented Storage Mechanisms}}
{
In this subsection, we introduce the task-oriented storage to collectively refers to both task caching approach, i.e., pre-storing programs/data for accelerated execution, and task buffering approach, i.e., dynamic task buffer management for multi-slot scheduling.
The task-oriented storage mechanisms aim to leverage storage resources to enhance MEC system flexibility and efficiency, while the two approaches have distinct mechanisms and optimization objectives.
}
Previous studies have investigated task caching and buffering from several different perspectives.
For instance, \cite{9076825} proposed the pre-storing scheme of the necessary data for processing tasks at MEC servers.
Then, in \cite{9419755}, the authors proposed proactive caching at the MEC server due to precedence dependencies among tasks.
To fully utilize the system resources, a joint server-device task processing and caching system was investigated \cite{9861697} considering parallel tasks.
The QoEs of users were optimized in \cite{10103637}, with the joint optimization of task caching, offloading, and resource allocation.

While these works effectively promoted task execution, they are fundamentally ill-suited for scenarios with stochastic task arrivals.
Consequently, for the stochastic and heterogeneous tasks inherent in generative AI and LAE applications, there is a critical necessity for more flexible buffering mechanisms capable of adaptive scheduling.
{It is worth noting that the task buffering differs from the task caching proposed in \cite{9076825, 9419755, 9861697, 10103637}. While the task caching pre-stores programs or data to accelerate known task execution, the task buffering manages the dynamic buffer queues to handle stochastic task arrivals across multiple time slots.}
Hence, task buffering serves as a crucial mechanism in MEC systems, decoupling task processing from immediate execution to enable flexible scheduling and robust handling of workload fluctuations. For instance, the  task buffer was introduced in a UAV-enabled MEC system \cite{10592015}, the long-term energy consumption was minimized with dynamic task arrival.
{Motivated by this, our work focuses on task buffering to enable flexible multi-slot scheduling under unpredictable task traffic conditions.
}

\begin{table*}[htbp]
\centering
\caption{Comparison of Related Works.}
\label{tab:comparison}
\renewcommand{\arraystretch}{1.3}
\resizebox{2\columnwidth}{!}{\begin{tabular}{lcccccc}
\toprule
\textbf{Reference} & \textbf{Movable antennas} & \textbf{MEC systems} & \textbf{Multi-time-slot resource allocation} & \textbf{MU mobility} & \textbf{Task-oriented storage} & \textbf{Task buffering} \\
\midrule
Refs. \cite{10354003, 11156108, 11261377}
& \checkmark & $\times$ & $\times$ & $\times$ & $\times$ & $\times$ \\
Ref. \cite{10654366}
& \checkmark & $\times$ & \checkmark & $\times$ & $\times$ & $\times$ \\
Ref. \cite{11363226}
& \checkmark & $\times$ & $\times$ & \checkmark & $\times$ & $\times$ \\
Refs. \cite{10528324, 11430678, 11017619, 11098592, 11301741, 11442647, 10971386}
& \checkmark & \checkmark & $\times$ & $\times$ & $\times$ & $\times$ \\
Ref. \cite{10620306}
& \checkmark & \checkmark & \checkmark & $\times$ & $\times$ & $\times$ \\
Refs. \cite{9076825, 9419755, 9861697}
& $\times$ & \checkmark & $\times$ & $\times$ & \checkmark & $\times$ \\
Ref. \cite{10103637}
& $\times$ & \checkmark & \checkmark & \checkmark & \checkmark & $\times$ \\
Ref. \cite{10592015}
& $\times$ & \checkmark & \checkmark & $\times$ & \checkmark & \checkmark \\
\cellcolor{mygray}{\textbf{\textbf{Our work}}}          & \cellcolor{mygray}{\checkmark} & \cellcolor{mygray}{\checkmark} & \cellcolor{mygray}{\checkmark} & \cellcolor{mygray}{\checkmark} & \cellcolor{mygray}{\checkmark} & \cellcolor{mygray}{\checkmark} \\
\bottomrule
\end{tabular}}
\end{table*}

\section{System Model}
\begin{figure}
    \centering
    \includegraphics[width=0.8\linewidth]{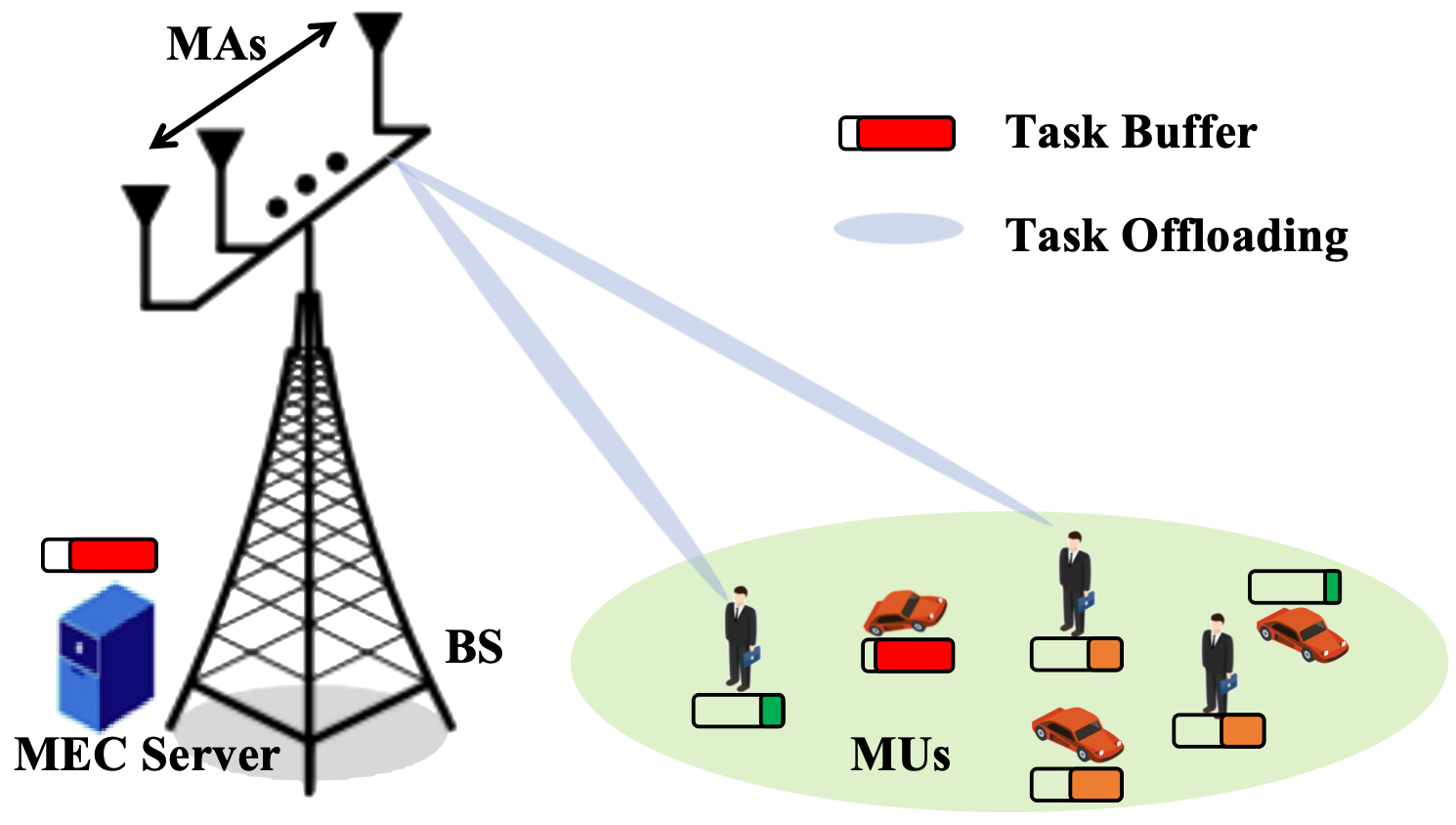}
    \caption{System model of the MA-enabled MEC system with task buffers.}
    \label{figsm}
\end{figure}

As presented in Fig. \ref{figsm}, the considered MEC system consists of one BS and $K$ MUs. The BS is enhanced by $M$ MAs, and is also equipped with an MEC server to provide computing services for the MUs. Each MU is equipped with a single FPA, and moving on the road.
The MEC server and the BS are connected with optical fiber. Hence, the data communication latency between the MEC server and the BS can be negligible.
The sets of MAs and MUs are denoted as $\mathcal{M} = \{1,\cdots, M\}$ and $\mathcal{K} = \{1,\cdots, K\}$, respectively. We consider a time interval, which has $T$ time slots, each with duration $\tau$ seconds. The set of the time slots is denoted as $\mathcal{T} = \{1,\cdots, T\}$.
The two-dimensional (2D) location of the BS is $[x_{\mathrm{BS}}, y_{\mathrm{BS}}]$, and the height of the BS is $H$. The 2D location of the $k$-th MU at time slot $t$ is given by $[x_{k}(t), y_{k}(t)]$.
{To improve readability, the main notations and the corresponding descriptions are given in Table \ref{tab:notations}.}

\begin{table}[h]
\centering
\footnotesize
\caption{Main Notations and Descriptions}
\label{tab:notations}
\begin{tabular}{c c}
\toprule
\textbf{Notations} & \textbf{Description} \\
\midrule
$k, K, \mathcal{K}$ & The index, number, and set of MUs \\
$m, M, \mathcal{M}$ & The index, number, and set of MAs \\
$t, T, \mathcal{T}$ & The index, number, and set of time slots \\
$\tau$ & The duration of each time slot \\
$x_m(t)$ & The position of the $m$-th MA at time slot $t$ \\
$\boldsymbol{x}(t)$ & The positions of all MAs at time slot $t$ \\
$d_{k,\text{BS}}(t)$ & The distance between the $k$-th MU and the BS \\
$h_{k\to\text{BS}}(t)$ & The channel coefficient between the $k$-th MU and the BS \\
$p_k$ & The transmit power of $k$-th MU \\
$\boldsymbol{w}_k(t)$ & The $k$-th column of $\mathbf{W}$ for the $k$-th MU \\
$r_k(t)$ & The offloading data rate of the $k$-th MU \\
$D_k(t)$ & The data size of the arrived task at the $k$-th MU \\
$f_k(t)$ & The scheduled computing resource of the $k$-th MU \\
$F_k^{\max}$ & The maximum computing capability of the $k$-th MU \\
$F(t)$ & The scheduled computing resource of the MEC server \\
$F^{\max}$ & The maximum computing capability of the MEC server \\
$D_k^{\text{local}}(t)$ & The locally computed data of the $k$-th MU \\
$D_k^{\text{off}}(t)$ & The offloaded data size of the $k$-th MU \\
$E_k(t)$ & The total energy consumption of the $k$-th MU \\
$C_k$ & The task buffer queue of the $k$-th MU \\
$C_{\text{MEC}}$ & The task buffer queue of the MEC server \\
\bottomrule
\end{tabular}
\end{table}

\subsection{Communications Model}
We consider that the channels among the BS and the MUs are determined by the propagation environment and the antenna position of the MA.
The position of the $m$-th MA at the BS is given by $x_m (t)$.
Hence, the positions of all MAs at time slot $t$ can be written as
$\boldsymbol{x}(t) = [x_1 (t), x_2 (t), \cdots, x_M (t) ]^{\mathsf{T}}$
\footnote{{
Following previous works \cite{10741192, 11154927, 11177504, 10693833, 11346858}, we focus on optimizing antenna positions to enhance uplink task offloading performance, thereby improving MEC efficiency and reducing energy consumption.
Hence, the mechanical energy consumption and latency overhead associated with MA movement are not explicitly modeled in this work.
Incorporating explicit MA movement constraints is an important direction for our future work.
}
}
\footnote{{
We consider one-dimensional linear MA array to explore the fundamental performance gains of MA-enabled MEC systems \cite{11177504, 10693833, 11346858}.
Linear MA arrays are also practically relevant for deployment along building edges, road sides, or UAV fuselages.
Extension to two-dimensional or area-based MA movement is an important direction for our future work.
}
}.
Similar to previous works \cite{10382559, 11374087}, the positions of the MAs satisfy $0 \leq x_1 (t) \leq x_2 (t) \leq \cdots \leq x_M (t) \leq x_{\max}$, where $x_{\max}$
is the maximum range of antennas.
We consider both the path loss and the small scale fading for the channel modeling.
Thus, the channel coefficient between the BS and the $k$-th MU is given by
\begin{equation}
\begin{split}
    \boldsymbol{h}_{k \rightarrow \mathrm{BS}} (t) = 
    \sqrt{\frac{\beta_0}{d_{k,\mathrm{BS}}^{2}(t)}}
    &\left(
        \sqrt{\frac{\kappa}{\kappa + 1}}
        \boldsymbol{h}_{k \rightarrow \mathrm{BS}}^{\mathsf{LoS}}(t)
    \right.\\
    &\quad \left. 
        + \sqrt{\frac{1}{\kappa + 1}}
        \boldsymbol{h}_{k \rightarrow \mathrm{BS}}^{\mathsf{NLoS}}(t)
    \right).
\end{split}
\end{equation}
Here, $\beta_0$ is the channel power at the reference distance of one meter, $d_{k,\mathrm{BS}}(t)$ represents the distance between the $k$-th MU and the BS, which is given by
\begin{equation}
d_{k,\mathrm{BS}}(t) = \sqrt{(x_{\mathrm{BS}}-x_k(t))^2+(y_{\mathrm{BS}}-y_k(t))^2+H^2}.
\end{equation}
And $\kappa$ denotes the Rician factor, $\boldsymbol{h}_{k \rightarrow \mathrm{BS}}^{\mathsf{NLoS}}(t)$ is the non-LoS (NLoS) component, which is a complex Gaussian random
vector, and each element has zero mean and unit covariance. The LoS component $\boldsymbol{h}_{k \rightarrow \mathrm{BS}}^{\mathsf{LoS}}(t)$ is given by
\begin{equation}
\begin{split}
    \boldsymbol{h}_{k \rightarrow \mathrm{BS}}^{\mathsf{LoS}} (t) = 
    &\left[
            \mathrm{e}^{\mathrm{j}\frac{2\pi x_1(t)}{\lambda}\cos \Phi_{k \rightarrow \mathrm{BS}}(t)},\right.\\
    &\quad\left.
            \cdots,
            \mathrm{e}^{\mathrm{j}\frac{2\pi x_M(t)}{\lambda}\cos \Phi_{k \rightarrow \mathrm{BS}}(t)}
    \right]^{\mathsf{T}},
\end{split}
\end{equation}
where $\Phi_{k \rightarrow \mathrm{BS}} (t)$ is the steering angle of the $k$-th MU towards the BS,
{which is given by
}
\begin{equation}
    \Phi_{k \rightarrow \mathrm{BS}} (t) = \arccos \left (
        \frac{x_{\mathrm{BS}}-x_k(t)}{d_{k,\mathrm{BS}}^{}(t)}
    \right ).
\end{equation}

Let $\boldsymbol{s}(t) = [s_1(t), s_2(t),\cdots, s_K(t)] $ denote the offloading signal of all MUs, and $p_k$ denote the transmit power of the $k$-th MU, respectively. Hence, the received signal at the BS is written as
\begin{equation}
    \boldsymbol{y}(t) = \sum_{k=1}^K \sqrt{p_k}\boldsymbol{h}_{k \rightarrow \mathrm{BS}} (t) s_k(t) + \boldsymbol{n}(t),
\end{equation}
where $\boldsymbol{n}(t) = [n_1(t), n_2(t), \cdots, n_M(t)]^{\mathsf{T}}$ denotes the noise vector, and $n_m \sim \mathcal{C}\mathcal{N}(0,\sigma^2)$ for $m \in \mathcal{M}$. The received beamforming matrix $\boldsymbol{W} \in \mathbb{C}^{M\times K}$ is introduced to estimate the received signal.
Hence, the achievable signal-to-interference-plus-noise ratio of the $k$-th MU is written as
\begin{equation}
    \gamma_k(t) = 
    \frac{p_k \vert \boldsymbol{w}^{\mathsf{H}}_k(t) \boldsymbol{h}_{k \rightarrow \mathrm{BS}} (t)\vert^2}
    {\sum_{i=1,i\neq k}^K p_i \vert \boldsymbol{w}^{\mathsf{H}}_k(t) \boldsymbol{h}_{i \rightarrow \mathrm{BS}} (t)\vert^2 + 
    \sigma^2 \Vert \boldsymbol{w}^{\mathsf{}}_k(t) \Vert^2}.
\end{equation}
Here, $\boldsymbol{w}^{\mathsf{}}_k(t)$ denotes the $k$-th column of $\boldsymbol{W} (t)$, corresponds for estimating the signal transmitted by the $k$-th MU, and the feasible region is given by
\begin{equation}
    \Vert \boldsymbol{w}^{\mathsf{}}_k(t) \Vert \leq 1,\
    \forall k \in \mathcal{K},\ \forall t \in \mathcal{T}.
\end{equation}
Hence, the offloading data rate of the $k$-th MU can be given by
$r_k (t) = B\log_2(1+\gamma_k(t))$,
where $B$ is the system bandwidth.

\subsection{Computing Model}
At the beginning of every time slot, each MU has a task that needs to be processed.
Let $D_k (t)$ denote the data size of the arrived task at the $k$-th MU in time slot $t$, which follows Poisson distribution with density $\rho_D$.
To ensure the computing requirement, the arrived tasks are clipped with a lower bound and an upper bound, i.e. $D_k (t) \in [D^{\min}, D^{\max}]$. 
Assume that the maximum computing capability of the $k$-th MU is $F_k^{\max}$, and the scheduled computing resource at time slot $t$ is $f_k (t) \leq F_k^{\max}$, the computed data of the $k$-th MU locally is given by
\begin{equation}
    D_k^{\mathrm{local}} (t) = \frac{f_k (t) \tau}{\xi},
\end{equation}
where $\xi$ is the required number of computing cycles per bit.
Accordingly, the energy consumption of the $k$-th MU for local computing can be expressed as
\begin{equation}
    E_k^{\mathrm{local}} (t) = \kappa_{\mathrm{comp}} f_k^3 (t) \tau,
\end{equation}
where $\kappa_{\mathrm{comp}}$ denotes the computing coefficient \cite{10233456, 11086503}.

Besides, the offloaded data size of the $k$-th MU to the BS can be given by
    $D_k^{\mathrm{off}} (t) = r_k(t) \tau$,
and the energy consumption for data offloading is $E_k^{\mathrm{off}} (t) = p_k \tau$. Therefore, the total energy consumption of the $k$-th MU at time slot $t$ can be given by
\begin{equation}
    E_k^{\mathrm{}} (t) = E_k^{\mathrm{local}} (t) + E_k^{\mathrm{off}} (t).
\end{equation}

Since the computing and offloading capability of MUs may not process all arrived task data, we assume that the task can be cached at the memory of the $k$-th MU. Let $\mathcal{C}_k$ denote the task buffer queue of the $k$-th MU, and the caching process can be expressed as
\begin{equation}\label{MUqueue}
    \mathcal{C}_k (t+1) = [\mathcal{C}_k (t) - D_k^{\mathrm{local}} (t) - D_k^{\mathrm{off}} (t)]^+ + D_k(t).
\end{equation}
Here, the operator $[ x ]^+ $ ensures that the result is never negative, i.e., $[ x ]^+ = \max (x,0)$.

The MEC server at the BS needs to process all the offloaded data. Let $F^{\max}$ denote the maximum computing capability of the MEC server, and the scheduled computing resource at time slot $t$ is $F(t) \leq F^{\max}$. The computed data can be expressed as
\begin{equation}
    D_{\mathrm{MEC}}(t)=\frac{F(t)\tau}{\xi}.
\end{equation}
Accordingly, the energy consumption for computing of the MEC server is $E_{\mathrm{MEC}}(t)=\kappa_{\mathrm{comp}} F^3 (t) \tau$. Similarly, the MEC server has a task buffer queue denoted as $\mathcal{C}_{\mathrm{MEC}}$, and the caching process of the MEC server can be given by
\begin{equation}\label{MECqueue}
    \mathcal{C}_{\mathrm{MEC}} (t+1) = [\mathcal{C}_{\mathrm{MEC}} (t) - D_{\mathrm{MEC}} (t)]^+ + D_{\mathrm{MEC}}^{\mathrm{off}}(t),
\end{equation}
where $D_{\mathrm{MEC}}^{\mathrm{off}}(t) = \sum_{k=1}^K D_k^{\mathrm{off}}(t)$ is the offloaded task data received by the MEC server at time slot $t$ \footnote{
{Within each time slot, the uplink task and MEC computation are executed in parallel. Specifically, while the BS is receiving offloaded task data from the MUs through uplink transmission, the MEC server simultaneously processes tasks from the task buffer queue. The parallel execution architecture is practically feasible since BSs are typically equipped with dedicated communication and computing modules that can operate concurrently. The buffer queue-enabled mechanism enables this pipeline processing, where the newly arrived offloaded data is stored in the task buffer queue at the MEC server and can be processed in the current or subsequent time slots, providing flexibility for multi-slot scheduling under bursty traffic conditions.
}
}.

\begin{definition}
The task buffer $\mathcal{C}_{\mathrm{MEC}}$ at the MEC server and $\mathcal{C}_{k}, \forall k \in \mathcal{K}$ at the MUs are strongly stable if the average queue length satisfies
\begin{equation}
    \lim _{T \rightarrow \infty} \frac{1}{T} \sum_{t=1}^{T} \mathbb{E}\left[\mathcal{C}_{\mathrm{MEC}}(t)\right]<\infty,
\end{equation}
\begin{equation}
    \lim _{T \rightarrow \infty} \frac{1}{T} \sum_{t=1}^{T} \mathbb{E}\left[\mathcal{C}_{k}(t)\right]<\infty,\quad \forall k \in \mathcal{K},
\end{equation}
where the expectations are taken with respect to the MEC system's random events, i.e., the arriving tasks, and the time varying channel conditions \cite{10592015}.
\end{definition}

\subsection{Problem Formulation}
In this paper, we aim to improve the energy efficiency of the MEC system enhanced by the MAs.
The total energy consumption for data processing is minimized, by jointly optimizing the receive beamforming $\boldsymbol{w} = \{ \boldsymbol{w}_k(t), \forall k, t \}$, the antenna position $\boldsymbol{x} = \{ x_m (t), \forall m, t \}$, and the computing resource scheduling $\boldsymbol{F} = \{ f_k(t), \forall k, t, F(t), \forall t \}$.
The problem is formulated as follows
\begin{subequations} { 
\begin{flalign}
 (\textbf{P1}):\ &\min_{\boldsymbol{w},\boldsymbol{x},\boldsymbol{F}} \quad  \sum_{t= 1}^{T} \left( \sum_{k= 1}^{K}E_{k}(t) + E_{\mathrm{MEC}}(t) \right)  \\
 {\rm{s.t.}}:  \quad 
 &\Vert \boldsymbol{w}^{\mathsf{}}_k(t) \Vert \leq 1,\
    \forall k \in \mathcal{K},\ \forall t \in \mathcal{T}, \label{p1a} \\
 & \{x_m(t)\}_{m=1}^M \in [0,x_{\max}],\ \forall t\in \mathcal{T}, \label{p1b} \\
 &{ x_{p} (t) - x_{q} (t)  \geq x_{\min}, \ \text{where }\{p, q\} \in \mathcal{M}} , \nonumber\\
 &\quad\quad\quad\quad\quad\quad\quad\quad\quad\quad\quad
 { p > q,\ \forall t\in \mathcal{T} },\label{p1c}\\
 & 0 \le f_{k}(t) \le F_{k}^{\text{max}},\ \forall k \in \mathcal{K},\ \forall t\in \mathcal{T}, \label{p1d} \\
 & 0 \le F(t) \le F_{}^{\max},\ \forall t\in \mathcal{T}, \label{p1e} \\
 & \lim _{T \rightarrow \infty} \frac{1}{T} \sum_{t=1}^{T} \mathbb{E}\left[\mathcal{C}_{\mathrm{MEC}}(t)\right]<\infty, \label{p1f} \\
 & \lim _{T \rightarrow \infty} \frac{1}{T} \sum_{t=1}^{T} \mathbb{E}\left[\mathcal{C}_{k}(t)\right]<\infty,\  \forall k \in \mathcal{K}, \label{p1g} 
\end{flalign}}\end{subequations}where $x_{\min}$ denotes the minimum distance between two antennas to avoid the coupling effect. 

\section{Problem Transformation and Solution}

\subsection{Problem Transformation}

First, to simplify the problem, the constraints of the antenna position \eqref{p1b} and \eqref{p1c} can be transformed through the following Proposition \ref{propositon2}.
\begin{proposition}
\label{propositon2}
    For any $m \in \mathcal{M}$, the position constraint of the $m$-th antenna can be given by
    \begin{equation}
        x_{m-1} (t) +x_{\min} \leq x_m (t)
        \leq x_{\max}-(M-m)x_{\min}.
        \label{p21}
    \end{equation}
    \begin{proof}
    Please refer to Appendix \ref{appprp1}.
    \end{proof}

\end{proposition}
Thus, (\textbf{P1}) can be rewritten as (\textbf{P2}), which is given by
\begin{subequations} { 
\begin{flalign}
 (\textbf{P2}):\ &\min_{\boldsymbol{w},\boldsymbol{x},\boldsymbol{F}} \quad  \sum_{t= 1}^{T} \left( \sum_{k= 1}^{K}E_{k}(t) + E_{\mathrm{MEC}}(t) \right)  \\
 {\rm{s.t.}}:  \quad &\eqref{p1a}, \eqref{p1d}-\eqref{p1g},\text{ and }\eqref{p21}.\nonumber
\end{flalign}}
\end{subequations}

Since problem (\textbf{P2}) and constraints \eqref{p1f} and \eqref{p1g} are time-coupled, which leads to the difficulty to solve problem (\textbf{P2}) through conventional optimization methods.
To decouple problem (\textbf{P2}) into each time slot for optimization, and maintain the stability of the task buffer queues of the MEC server and the MUs, we first introduce the Lyapunov function $\mathcal{L}(\mathcal{C}(t))$ and conditional Lyapunov drift $\nabla \mathcal{L}(\mathcal{C}(t))$ \cite{10592015}, which are given by
\begin{equation}
    \mathcal{L}(\mathcal{C}(t)) = \frac{1}{2}
    \left \{
        (\mathcal{C}_{\mathrm{MEC}}(t))^2 + 
        \sum_{k=1}^K (\mathcal{C}_{k}(t))^2
    \right \},
\end{equation}

\begin{align}
&\nabla\mathcal{L}(\mathcal{C}(t)) = \mathbb{E}\left[ \mathcal{L}(\mathcal{C}(t+1)) - \mathcal{L}(\mathcal{C}(t)) \mid \mathcal{C}(t) \right] \nonumber \\
&= \frac{1}{2} \mathbb{E}\left[ \left( \max\{(\mathcal{C}_{\mathrm{MEC}}(t) - {D}_{\mathrm{MEC}}(t)), 0\} + {D}_{\mathrm{MEC}}^{\mathrm{off}}(t) \right)^2 \;\Bigg|\; \mathcal{C}(t) \right] \nonumber \\
&\quad + \frac{1}{2} \sum_{k=1}^{K} \mathbb{E}\left[ \left( \max\{(\mathcal{C}_k(t) - {D}_k^{\mathrm{out}}(t)), 0\} + {D}_k(t) \right)^2 \;\Bigg|\; \mathcal{C}(t) \right] \nonumber \\
&\quad - \frac{1}{2}(\mathcal{C}_{\mathrm{MEC}}(t))^2 - \frac{1}{2}\sum_{k=1}^{K}(\mathcal{C}_k(t))^2,
\label{drift}
\end{align}
where $\mathcal{C}(t)= \{ \mathcal{C}_{\mathrm{MEC}}(t), \mathcal{C}_{k}(t), \forall k \}$ is the total buffer queue, and $D_{k}^{\mathrm{out}}(t) = D_{k}^{\mathrm{local}}(t) + D_{k}^{\mathrm{off}}(t)$ is the processed data of the $k$-th MU.
The following Lemma \ref{le1} shows that the expected conditional Lyapunov drift $\nabla \mathcal{L}(\mathcal{C}(t))$ is upper bounded.
\begin{lemma}\label{le1}
    For any receive beamforming $\boldsymbol{w}$, antenna position $\boldsymbol{x}$, computing resource scheduling $\boldsymbol{F}$, and task buffer queue $\mathcal{C}$ satisfied constraints \eqref{p1a}, \eqref{p1d}-\eqref{p1g} and \eqref{p21}, the upper bound of the conditional Lyapunov drift $\nabla \mathcal{L}(\mathcal{C}(t))$ is given by
\begin{equation}
\begin{split}
    \nabla \mathcal{L}(\mathcal{C}(t)) \leq&\
    \mathcal{C}_{\mathrm{MEC}} (t) \mathbb{E}[D_{\mathrm{MEC}}^{\mathrm{off}}(t)-D_{\mathrm{MEC}}(t)\vert \mathcal{C}(t)]\\
    &+ \sum_{k=1}^K \mathcal{C}_{k} (t) \mathbb{E}[D_{k}^{\mathrm{}}(t)-D_{k}^{\mathrm{out}}(t)\vert \mathcal{C}(t)]
    + \Tilde{\mathcal{C}}.
\end{split}
\label{driftlower}
\end{equation}

\begin{proof}
Please refer to Appendix \ref{apple1}.
\end{proof}
\end{lemma}

We introduce the drift-plus-penalty function of each time slot to balance the buffer queue optimization and the energy optimization, which is given by
\begin{equation}
    \Phi(\mathcal{C}(t)) = 
    \nabla \mathcal{L}(\mathcal{C}(t)) +
    \Gamma \mathbb{E}\left[\sum_{k= 1}^{K}E_{k}(t) + E_{\mathrm{MEC}}(t)\vert \mathcal{C}(t)\right],
\end{equation}
where $\Gamma$ is the balance coefficient between the buffer queue optimization and the energy optimization. Based on Lemma \ref{le1}, the drift-plus-penalty $\Phi(\mathcal{C}(t))$ is also upper bounded, which is given by
\begin{equation}
\begin{split}
    \Phi(\mathcal{C}(t)) \leq &\
    \mathcal{C}_{\mathrm{MEC}} (t) \mathbb{E}[D_{\mathrm{MEC}}^{\mathrm{off}}(t)-D_{\mathrm{MEC}}(t)\vert \mathcal{C}(t)]\\
    &+ \sum_{k=1}^K \mathcal{C}_{k} (t) \mathbb{E}[D_{k}^{\mathrm{}}(t)-D_{k}^{\mathrm{out}}(t)\vert \mathcal{C}(t)] \\& +  \Tilde{\mathcal{C}} +
    \Gamma \mathbb{E}\left[\sum_{k= 1}^{K}E_{k}(t) + E_{\mathrm{MEC}}(t)\vert \mathcal{C}(t)\right].
\end{split}
\label{dppupper}
\end{equation}
Problem (\textbf{P2}) can be transformed when the right hand side of \eqref{dppupper} is minimized, and the transformed problem is given by
\begin{subequations} { 
\begin{flalign}
 (\textbf{P3}):\ &\min_{\boldsymbol{w},\boldsymbol{x},\boldsymbol{F}} \quad  \mathcal{C}_{\mathrm{MEC}} (t) \mathbb{E}[D_{\mathrm{MEC}}^{\mathrm{off}}(t)-D_{\mathrm{MEC}}(t)\vert \mathcal{C}(t)] \nonumber\\
    &\quad\quad\quad\quad + \sum_{k=1}^K \mathcal{C}_{k} (t) \mathbb{E}[D_{k}^{\mathrm{}}(t)-D_{k}^{\mathrm{out}}(t)\vert \mathcal{C}(t)] \nonumber\\&\quad\quad\quad\quad +  
    \Gamma \left(\sum_{k= 1}^{K}E_{k}(t) + E_{\mathrm{MEC}}(t)\right) + \Tilde{\mathcal{C}} \\
 {\rm{s.t.}}:  \quad & \eqref{p1a}, \eqref{p1d},\eqref{p1e},\text{ and }\eqref{p21} \nonumber.
\end{flalign}}\end{subequations}
\begin{rmk}
    According to the Lyapunov-based decoupling, the current task buffer queues $\mathcal{C}_{\mathrm{MEC}} (t)$ and $\mathcal{C}_{k} (t), \forall k$ are constant during the corresponding time slot.
    Meanwhile, $\Tilde{\mathcal{C}}$ is a constant, and can be omitted in the objective function.
    Moreover, the task arrival $\mathcal{D}_k(t)$ are observed at the beginning of slot $t$,
    and the energy consumption of MUs and the MEC server, as well as the process task data are deterministic functions of the optimization variables.
    Problem (\textbf{P3}) is equivalent to the following deterministic optimization problem (\textbf{P3-1}).
\end{rmk}
\begin{subequations}{ 
\begin{flalign}
 (\textbf{P3-1}):\ &\min_{\boldsymbol{w},\boldsymbol{x},\boldsymbol{F}} \quad  \mathcal{C}_{\mathrm{MEC}} (t) \left(D_{\mathrm{MEC}}^{\mathrm{off}}(t)-D_{\mathrm{MEC}}(t)\right) \nonumber\\
    &\quad\quad\quad\quad + \sum_{k=1}^K \mathcal{C}_{k} (t) \left(D_{k}^{\mathrm{}}(t)-D_{k}^{\mathrm{out}}(t)\right)\nonumber\\&\quad\quad\quad\quad +  
    \Gamma \left(\sum_{k= 1}^{K}E_{k}(t) + E_{\mathrm{MEC}}(t)\right) \label{objp31} \\
 {\rm{s.t.}}:  \quad & \eqref{p1a}, \eqref{p1d},\eqref{p1e},\text{ and }\eqref{p21} \nonumber.
\end{flalign}}
\end{subequations}

Problem (\textbf{P3-1}) is still non-convex, we utilize an efficient AO algorithm to solve (\textbf{P3-1}), by decomposing it into three subproblems, which are the optimization of receive beamforming, the optimization of the antenna position, and the optimization of computing resource allocation, respectively.

\subsection{Receive Beamforming Optimization}
Given the antenna position $\boldsymbol{x}$ and the computing resource allocation $\boldsymbol{F}$, 
the objective function of problem (\textbf{P3}) can be simplified as
\begin{equation}
\begin{split}
    &\mathcal{C}_{\mathrm{MEC}} (t) (D_{\mathrm{MEC}}^{\mathrm{off}}(t)-D_{\mathrm{MEC}}(t)) \\
    &\quad\quad\quad\quad + \sum_{k=1}^K \mathcal{C}_{k} (t)( D_{k}^{\mathrm{}}(t)-D_{k}^{\mathrm{out}}(t))\\
    &\quad\quad\quad\quad +  
    \Gamma \left(\sum_{k= 1}^{K}E_{k}(t) + E_{\mathrm{MEC}}(t)\right) \\
    &=\sum_{k=1}^K \Upsilon_k(t) \log_2 (1+\gamma_k(t)) + \Omega (t),
\end{split}
\end{equation}
where $\Upsilon_k(t)$
and $\Omega(t)$ are two constants, and given by
\begin{equation}
    \Upsilon_k(t) = \tau B (\mathcal{C}_{\mathrm{MEC}} (t) - \mathcal{C}_{k} (t)),
\end{equation}
\begin{equation}
\begin{split}
    \Omega(t) =&\ \sum_{k=1}^K \mathcal{C}_{k} (t) (D_k(t)-D_k^{\mathrm{local}}(t))-\mathcal{C}_{\mathrm{MEC}} (t)D_{\mathrm{MEC}}(t) \\&+\Gamma \left(\sum_{k= 1}^{K}E_{k}(t) + E_{\mathrm{MEC}}(t)\right).
\end{split}
\end{equation}
Since $\Omega(t)$ is not related to the optimization of receive beamforming, it can be omitted.
Hence, problem (\textbf{P3}) can be simplified as problem (\textbf{P4}), which is given by
\begin{subequations} {  
\begin{flalign}
 (\textbf{P4}):\ &\min_{\boldsymbol{w}} \quad  \sum_{k=1}^K \Upsilon_k(t) \log_2 (1+\gamma_k(t))\\
 {\rm{s.t.}}:  \quad & \eqref{p1a}. \nonumber 
\end{flalign}}\end{subequations}
\begin{rmk}
\label{upsilon}
    The coefficient $\Upsilon_k(t)$ has a physical interpretation for offloading decisions.
    When $\Upsilon_k(t) > 0$, i.e., $\mathcal{C}_{\mathrm{MEC}} (t) > \mathcal{C}_{k} (t)$, the data size of the MEC server's task buffer is larger than the local task buffer of the $k$-th MU.
    In this case, further offloading would exacerbate the congestion at the MEC server while depleting local buffer of the $k$-th MU. From a buffer balance perspective, the decision is to suspend the offloading transmission from the $k$-th MU {scheduled by the MEC server}, i.e., the objective function in problem (\textbf{P4}) naturally drives $\boldsymbol{w}_k \to \boldsymbol{0}$. 
    When $\Upsilon_k(t) \leq 0$, i.e., $\mathcal{C}_{\mathrm{MEC}} (t) \leq \mathcal{C}_{k} (t)$, the $k$-th MU has more buffered tasks than the MEC server. In this case, offloading is beneficial for balancing the buffer size with the reduced overall Lyapunov drift. Hence, the receive beamforming for the $k$-th MU should be optimized to maximize the offloading data rate.
\end{rmk}
Following Remark \ref{upsilon},
when $\Upsilon_k(t) \leq 0$, problem (\textbf{P4}) is equivalent to the problem (\textbf{P4-1}), which is given by
\begin{subequations} {  
\begin{flalign}
 (\textbf{P4-1}):\ &\max_{\boldsymbol{w}} \ \sum_{k\in \Tilde{\mathcal{K}}} \Tilde{\Upsilon}_k(t) \log_2 (1+\gamma_k(t)) \\
 {\rm{s.t.}}:  \quad & \eqref{p1a}, \nonumber 
\end{flalign}}\end{subequations}
where $\Tilde{\mathcal{K}} = \{k \vert \Upsilon_k(t) \leq 0 \}$ is the set of MUs for whom offloading is beneficial, and $\Tilde{\Upsilon}_k(t) = - \Upsilon_k(t)$.
Let $\boldsymbol{W}_k = \boldsymbol{w}_k\boldsymbol{w}_k^{\mathsf{H}}$, and $\boldsymbol{W}_k$ satisfies  $\mathsf{rank}(\boldsymbol{W}_k) \leq 1$ and $\boldsymbol{W}_k \succeq 0$. By substituting $\boldsymbol{W}_k$ into problem (\textbf{P4-1}), the problem can be reformulated by semi-definite relaxation (SDR) as
\begin{subequations} {  
\begin{flalign}
 (\textbf{P4-2}):\ &\max_{\boldsymbol{W}_k} \  \sum_{k\in \Tilde{\mathcal{K}}} \Tilde{\Upsilon}_k(t) \Tilde{r}_k(t) \\
 {\rm{s.t.}}:  \quad & \mathsf{tr}(\boldsymbol{W}_k(t)) \leq 1, \label{p3-2-1}\\
 & \boldsymbol{W}_k \succeq 0, \label{p3-2-2}\\
 & \mathsf{rank}(\boldsymbol{W}_k) \leq 1, \label{p3-2-3}
\end{flalign}}\end{subequations}
where $\tilde{r}_k(t)$ is given by
\begin{equation}
    \tilde{r}_k(t) = 
    \log_2 \left (1+
        \frac{\mathsf{tr}\left( p_k \boldsymbol{H}_{k \rightarrow \mathrm{BS}}(t) \boldsymbol{W}_k(t) \right)}{
        I_k(t)
        + \mathsf{tr} \left( \sigma^2 \boldsymbol{W}_k(t)
        \right)
        }
    \right ).
    \label{transformedrate}
\end{equation}
{Here, $\boldsymbol{H}_{k \rightarrow \mathrm{BS}}(t) = \boldsymbol{h}_{k \rightarrow \mathrm{BS}}(t) \boldsymbol{h}_{k \rightarrow \mathrm{BS}}^{\mathsf{H}}(t)$,
and $I_k(t)$ is the interference caused by other MUs, which can be expressed as
\begin{equation}
    I_k(t) = \sum_{i\in \Tilde{\mathcal{K}}, i \neq k}
        \mathsf{tr}\left( p_i \boldsymbol{H}_{i \rightarrow \mathrm{BS}}(t)\boldsymbol{W}_k(t) \right).
\end{equation}
\eqref{transformedrate} is still non-convex, which can be further written by a concave-minus-concave form as
\begin{equation}
\begin{split}
    \tilde{r}_k(t) =\ &\log_2 \left ( \sum_{i\in \tilde{\mathcal{K}}}
        \mathsf{tr}\left( p_i \boldsymbol{H}_{i \rightarrow \mathrm{BS}}(t)\boldsymbol{W}_k(t) \right) + \mathsf{tr} \left( \sigma^2 \boldsymbol{W}_k(t) \right )
        \right)\\
        &- \log_2 \left( I_k(t)
        + \mathsf{tr} \left( \sigma^2 \boldsymbol{W}_k(t)
        \right) \right).
\end{split}
\label{logmlog}
\end{equation}
By introducing Taylor expansion on the second concave term in \ref{logmlog}, we can obtain the lower bound of $\tilde{r}_k(t)$ in the $(l)$-th iteration of successive convex approximation (SCA) as
\begin{equation}
\begin{split}
    &\tilde{r}_k(t) \geq \log_2 \left ( \sum_{i\in \tilde{\mathcal{K}}}
        \mathsf{tr}\left( p_i \boldsymbol{H}_{i \rightarrow \mathrm{BS}}(t)\boldsymbol{W}_k(t) \right) + \mathsf{tr} \left( \sigma^2 \boldsymbol{W}_k(t) \right )
        \right) \\
    &- \log_2 \left( I_k^{(l)}(t)
        + \mathsf{tr} \left( \sigma^2 \boldsymbol{W}_k^{(l)}(t)
        \right) \right)\\
    &-  \sum_{i\in \tilde{\mathcal{K}}, i \neq k} \mathsf{tr} \left (
        \frac{\log_2 (\mathrm{e})\left ( p_i \boldsymbol{H}_{i \rightarrow \mathrm{BS}}(t) \left( \boldsymbol{W}_k(t) - \boldsymbol{W}_k^{(l)}(t)
        \right)
        \right )
        }{I_k^{(l)}(t)
        + \mathsf{tr} \left( \sigma^2 \boldsymbol{W}_k^{(l)}(t) \right)}
    \right )\\
    &-  \mathsf{tr} \left (
        \frac{\log_2 (\mathrm{e})\left (  \sigma^2 \left( \boldsymbol{W}_k(t) - \boldsymbol{W}_k^{(l)}(t)
        \right)
        \right )
        }{I_k^{(l)}(t)
        + \mathsf{tr} \left( \sigma^2 \boldsymbol{W}_k^{(l)}(t) \right)}
    \right )\\
    &\triangleq \check{r}_k(t),
\end{split}
\end{equation}
where $I_k^{(l)}(t) = \sum_{i\in \tilde{\mathcal{K}}, i \neq k}
        \mathsf{tr}\left( p_i \boldsymbol{H}_{i \rightarrow \mathrm{BS}}(t)\boldsymbol{W}_k^{(l)}(t) \right)$.
Accordingly, problem (\textbf{P4-2}) can be written as
\begin{subequations} {  
\begin{flalign}
 (\textbf{P4-2}.l):\ &\max_{\boldsymbol{W}_k} \ \sum_{k\in \tilde{\mathcal{K}}} \tilde{\Upsilon}_k(t) \check{r}_k(t) \\
 {\rm{s.t.}}:  \quad & \eqref{p3-2-1}, \eqref{p3-2-2}
\end{flalign}}\end{subequations}}

\begin{rmk}
    Problem \textbf{(P4-2)}.$l$ is convex, which can be efficiently solved using standard convex solvers, such as CVX.
Denoting the solution to this problem as $\boldsymbol{W}_k^*$, we observe that if $\mathsf{rank}\left(\boldsymbol{W}_k^*\right) \leq 1, \forall k$, the SDR is tight and the solutions are optimal for problem \textbf{(P4-1)}. While in cases that the rank exceeds one, we can leverage the rank-one reconstruction techniques like Gaussian randomization
\cite{9916163, 11373590}.
\end{rmk}

\subsection{Antenna Position Optimization}
Given the receive beamforming $\boldsymbol{w}$ and the computing resource allocation $\boldsymbol{F}$, similar to the optimization of receive beamforming, 
the objective function of problem (\textbf{P3}) can be simplified as
\begin{equation}
     (\textbf{P5}):\ \max_{\boldsymbol{x}} \   \sum_{k\in \Tilde{\mathcal{K}}} \Tilde{\Upsilon}_k(t) \log_2 (1+\gamma_k(t)) \  \
 {\rm{s.t.:}}  \  \eqref{p21}. 
\end{equation}

Noting that during the antenna position optimization, we focus primarily on the LoS component to model the interplay between array geometry and reception gain.
The reason is that optimizing antenna positions directly controls the phase relationships among array elements and the resulting coherent combining gain, both of which are predominantly determined by the LoS component.
Conversely, the NLoS component behaves as random fast fading, lacking a stable or analytically solvable geometric mapping to position variables.
Therefore,
optimizing positions based solely on the LoS component allows for the extraction of the dominant controllable factors in array layout design and clearly reveals the intrinsic relationship between element positions and array gain.
restricting the optimization to the LoS component isolates the dominant controllable terms in array design and elucidates the fundamental link between element placement and array gain.

Hence, we first rewrite $p_k \vert \boldsymbol{w}^{\mathsf{H}}_k(t) \boldsymbol{h}_{k \rightarrow \mathrm{BS}} (t) \vert^2$ as
\begin{equation}
\begin{split}
    &p_k \vert \boldsymbol{w}^{\mathsf{H}}_k(t) \boldsymbol{h}_{k \rightarrow \mathrm{BS}} (t) \vert^2 \approx \varphi_k(t) \left (
    \boldsymbol{a}_k^{\mathsf{T}} (t) \boldsymbol{U}_k^{(1)}  (t) \boldsymbol{a}_k^{\mathsf{}} (t) \right.
    \\& \left. \quad\quad+\boldsymbol{b}_k^{\mathsf{T}} (t) \boldsymbol{U}_k^{(1)}  (t) \boldsymbol{b}_k^{\mathsf{}} (t)
    +2\boldsymbol{a}_k^{\mathsf{T}} (t) \boldsymbol{U}_k^{(2)}  (t) \boldsymbol{b}_k^{\mathsf{}} (t) \right)
    \\ &\triangleq
    \Psi_k (\boldsymbol{a}_k(t), \boldsymbol{b}_k(t);
    \boldsymbol{w}_k(t)),
\end{split}
\end{equation}
where 
\begin{equation}
    \varphi_k(t) =  {\frac{p_k\beta_0\kappa}{d_{k,\mathrm{BS}}^{2}(t)(\kappa + 1)}},
\end{equation}
$\boldsymbol{w}_k (t) = \boldsymbol{u}_k (t) + \mathrm{j}\boldsymbol{v}_k (t)$, $\boldsymbol{U}_k^{(1)}  (t) = \boldsymbol{u}_k (t) \boldsymbol{u}_k^{\mathsf{T}} (t) + \boldsymbol{v}_k (t) \boldsymbol{v}_k^{\mathsf{T}} (t)$, and $\boldsymbol{U}_k^{(2)} (t)= \boldsymbol{u}_k (t) \boldsymbol{v}_k^{\mathsf{T}} (t) - \boldsymbol{v}_k (t) \boldsymbol{u}_k^{\mathsf{T}} (t)$. $\boldsymbol{a}_k(t)$ and $\boldsymbol{b}_k(t)$ are given by
\begin{equation}
    \boldsymbol{a}_k(t) = \left [ 
    a_k^1(t), a_k^2 (t), \cdots, a_k^M (t)
    \right ]^{\mathsf{T}},
\end{equation}
\begin{equation}
    \boldsymbol{b}_k(t) = \left [ 
    b_k^1(t), b_k^2 (t), \cdots, b_k^M (t)
    \right ]^{\mathsf{T}},
\end{equation}
where
\begin{equation}
    a_k^m(t) = \cos \left (
        \frac{2\pi x_m(t)}{\lambda}\cos \Phi_{k \rightarrow \mathrm{BS}}(t)
    \right ),
\end{equation}
\begin{equation}
    b_k^m(t) = \sin \left (
        \frac{2\pi x_m(t)}{\lambda}\cos \Phi_{k \rightarrow \mathrm{BS}}(t)
    \right ).
\end{equation}
Similarly, $p_i \vert \boldsymbol{w}^{\mathsf{H}}_k(t) \boldsymbol{h}_{i \rightarrow \mathrm{BS}} (t)\vert^2, i\neq k$ is rewritten as
\begin{equation}
\begin{split}
    &p_i \vert \boldsymbol{w}^{\mathsf{H}}_k(t) \boldsymbol{h}_{i \rightarrow \mathrm{BS}} (t) \vert^2 \approx \varphi_i(t) \left (
    \boldsymbol{a}_i^{\mathsf{T}} (t) \boldsymbol{U}_k^{(1)}  (t) \boldsymbol{a}_i^{\mathsf{}} (t) \right.
    \\& \left. \quad\quad+\boldsymbol{b}_i^{\mathsf{T}} (t) \boldsymbol{U}_k^{(1)}  (t) \boldsymbol{b}_i^{\mathsf{}} (t)
    +2\boldsymbol{a}_i^{\mathsf{T}} (t) \boldsymbol{U}_k^{(2)}  (t) \boldsymbol{b}_i^{\mathsf{}} (t) \right)
    \\ &\triangleq
    \Psi_i (\boldsymbol{a}_i(t), \boldsymbol{b}_i(t);
    \boldsymbol{w}_k(t)),\ i \neq k.
\end{split}
\end{equation}

As a result, problem (\textbf{P5}) can be rewritten as
problem (\textbf{P5-1}), which is given by 
\begin{subequations} {  
\begin{flalign}
 (\textbf{P5-1}):\ &\max_{\boldsymbol{x}} \quad  \Tilde{\Psi}\left( \boldsymbol{a}_i(t), \boldsymbol{b}_i(t) \vert_{i\in \Tilde{\mathcal{K}}} \right ) \\
 {\rm{s.t.}}:  \quad & \eqref{p21}. \nonumber 
\end{flalign}}\end{subequations}
The objective function of problem (\textbf{P5-1}) is defined by \eqref{oj45},
\begin{figure*}[!t]
\vspace*{-\baselineskip} 
\vspace*{1.5mm}
    \begin{equation}{\small
    \begin{split}
    \Tilde{\Psi}\left( \boldsymbol{a}_i(t), \boldsymbol{b}_i(t) \vert_{i\in \Tilde{\mathcal{K}}} \right )
\triangleq
  \sum_{k\in \Tilde{\mathcal{K}}} \Tilde{\Upsilon}_k(t) \left( \log_2 \left( \sum_{i\in \Tilde{\mathcal{K}}}  \Psi_i (\boldsymbol{a}_i(t), \boldsymbol{b}_i(t);\boldsymbol{w}_k(t) ) + \Tilde{w}_k(t) \right) 
  \right.
 -\left. \log_2 \left( \sum_{i\in \Tilde{\mathcal{K}}, i \neq k} \Psi_i (\boldsymbol{a}_i(t), \boldsymbol{b}_i(t);\boldsymbol{w}_k(t) ) + \Tilde{w}_k(t) \right ) \right),
\end{split}
\label{oj45}}
    \end{equation}
\hrulefill
\end{figure*}
where $\Tilde{w}_k(t) = \sigma^2 \vert \boldsymbol{w}^{\mathsf{}}_k(t) \vert^2$. Problem (\textbf{P5-1}) is still non-convex and difficult to be solved. Hence, we introduce the projected gradient ascent (PGA) method to find a local optimal solution to problem (\textbf{P5-1}) \cite{10416363, 11156108}. Two steps are iterated in the PGA method, where the first step corresponds to the gradient update, and the second step represents the projection update.

Specifically, the gradient update of the $j+1$-th iteration of the PGA is given by
\begin{equation}
    \boldsymbol{x}^{j+1}(t) = 
    \boldsymbol{x}^{j}(t)
    + \varrho \nabla_{\boldsymbol{x}^{j}(t)}
    \Tilde{\Psi}\left( \boldsymbol{a}_i(t), \boldsymbol{b}_i(t) \vert_{i\in \Tilde{\mathcal{K}}} \right ),
    \label{gd}
\end{equation}
where $\varrho$ denotes the step size for gradient ascent. The closed-form expression of the gradient $\nabla_{\boldsymbol{x}^{j}(t)}
    \Tilde{\Psi}\left( \boldsymbol{a}_i(t), \boldsymbol{b}_i(t) \vert_{i\in \Tilde{\mathcal{K}}} \right )$ is given by the following proposition.

\begin{proposition}
\label{pro2pga}
    The closed-form expression of the gradient $\nabla_{\boldsymbol{x}^{j}(t)}
    \Tilde{\Psi}\left( \boldsymbol{a}_i(t), \boldsymbol{b}_i(t) \vert_{i\in \Tilde{\mathcal{K}}} \right )$ is given by
    \begin{equation}
    \begin{split}
    &\nabla_{\boldsymbol{x}^{j}(t)}
    \Tilde{\Psi}\left( \boldsymbol{a}_i(t), \boldsymbol{b}_i (t) \vert_{i\in \Tilde{\mathcal{K}}} \right )
    \\&=
    \sum_{k\in \Tilde{\mathcal{K}}} \Tilde{\Upsilon}_k \left(   \frac{\sum_{i\in \Tilde{\mathcal{K}}} \Tilde{\boldsymbol{X}}_i (\boldsymbol{a}_i, \boldsymbol{b}_i;\boldsymbol{w}_k )}{\ln(2)\left( \sum_{i\in \Tilde{\mathcal{K}}}  \Psi_i (\boldsymbol{a}_i, \boldsymbol{b}_i;\boldsymbol{w}_k ) + \Tilde{w}_k \right)} \right. \\ &\quad\quad\quad\quad \left.
 \ \ -  \frac{ \sum_{i\in \Tilde{\mathcal{K}}, i \neq k} \Tilde{\boldsymbol{X}}_i (\boldsymbol{a}_i, \boldsymbol{b}_i;\boldsymbol{w}_k ) }{\ln(2) \left( \sum_{i\in \Tilde{\mathcal{K}}, i \neq k}  \Psi_i (\boldsymbol{a}_i, \boldsymbol{b}_i;\boldsymbol{w}_k ) + \Tilde{w}_k \right ) }
 \right).
\end{split}
\label{pro1eq1}
\end{equation}
\begin{proof}
Please refer to Appendix \ref{appgpa}.
\end{proof}
\end{proposition}

After the gradient update by \eqref{gd} in the first step, the projection update is introduced to check the feasibility of the gradient update.
Accordingly, we update $\boldsymbol{x}^j (t)$ of the $j$-th iteration after gradient update by
\begin{equation}
    \boldsymbol{x}^j (t) \leftarrow
    \mathcal{P} (\boldsymbol{x}^j (t); x_{\min}, x_{\max}),
\end{equation}
where $\mathcal{P} (\boldsymbol{x}^j (t); x_{\min}, x_{\max})$ is the projection function, and each antenna is updated by the index order of $\mathcal{M}$, and the projection update of the $m$-th antenna is given by
\begin{equation}
\begin{split}
        x_m^j \leftarrow
    \max \Big(&x_{m-1}^j(t) + x_{\min}, \\
    &\min (x_m^j(t), x_{\max}-(M-m)x_{\min
    }) \Big).
\end{split}
\end{equation}

\subsection{Computing Resource Allocation}
Given the receive beamforming $\boldsymbol{w}$ and the antenna position $\boldsymbol{x}$, 
the objective function of problem (\textbf{P3}) can be simplified as
\begin{subequations} {  
\begin{flalign}
 (\textbf{P6}):\ &\min_{\boldsymbol{F}} \quad  \Bar{A}\sum_{k=1}^K f^3_k (t) - \frac{\tau}{\xi}\sum_{k=1}^K \mathcal{C}_k(t)f_k(t) \nonumber \\ &\quad\quad\quad+
\Bar{A} F^3(t) - \frac{\tau \mathcal{C}_{\mathrm{MEC}(t)}}{\xi} F(t) + \Bar{B}(t) \\
 {\rm{s.t.}}:  \quad & \eqref{p1d}{\text{ and }}\eqref{p1e},\nonumber 
\end{flalign}}\end{subequations}
where $\Bar{A} = \tau \kappa_{\mathrm{comp}} \Gamma$, and $\Bar{B}(t) = \mathcal{C}_{\mathrm{MEC}}(t)D_{\mathrm{MEC}}^{\mathrm{off}}(t) + \sum_{k=1}^K\mathcal{C}_k(t)(D_k(t)-D_k^{\mathrm{off}}(t))$, for simplicity.
Since the objective function of problem (\textbf{P6}) and all constraints are convex, problem (\textbf{P6}) is a convex problem. Hence, it can be solved by the standard methods, such as interior point method.

\subsection{Algorithm Analysis}

\begin{algorithm}
	\caption{Lyapunov-guided AO Algorithm for Solving (\textbf{P1})}
	\label{algoverall}
	\begin{algorithmic}[1]
		\REQUIRE Initial feasible solution $\{{\boldsymbol{w}}^{0}(1),{\boldsymbol{x}}^{0}(1),{\boldsymbol{F}}^{0}(1)\}$, $\mathcal{C}_k(1)$, and $\mathcal{C}_{\rm MEC}(1)$;\\
		\textbf{Initialize:} the precision threshold $\epsilon$;\\
		\FOR{$t=1$ to $T$}
		\STATE Set $\ell=0$ and initialize $\{{\boldsymbol{w}}^{0}(t),{\boldsymbol{x}}^{0}(t),{\boldsymbol{F}}^{0}(t)\}$;\\
		\STATE Calculate the objective value $\Phi^{0}(\mathcal{C}(t))$ of problem (\textbf{P3-1});\\
		\REPEAT
		\STATE Given ${\boldsymbol{x}}^{\ell}(t)$ and ${\boldsymbol{F}}^{\ell}(t)$, solve problem (\textbf{P4-2}.$l$) to obtain ${\boldsymbol{w}}^{\ell+1}(t)$;\\
		\STATE Given ${\boldsymbol{w}}^{\ell+1}(t)$ and ${\boldsymbol{F}}^{\ell}(t)$, solve problem (\textbf{P5-1}) by the PGA method to obtain ${\boldsymbol{x}}^{\ell+1}(t)$;\\
		\STATE Given ${\boldsymbol{w}}^{\ell+1}(t)$ and ${\boldsymbol{x}}^{\ell+1}(t)$, solve problem (\textbf{P6}) to obtain ${\boldsymbol{F}}^{\ell+1}(t)$;\\
		\STATE Calculate the objective value $\Phi^{\ell+1}(\mathcal{C}(t))$ of problem (\textbf{P3-1});\\
		\STATE Set $\ell=\ell+1$;\\
		\UNTIL{$|\Phi^{\ell}(\mathcal{C}(t))-\Phi^{\ell-1}(\mathcal{C}(t))|<\epsilon$;}\\
		\STATE Set ${\boldsymbol{w}}(t)={\boldsymbol{w}}^{\ell}(t)$, ${\boldsymbol{x}}(t)={\boldsymbol{x}}^{\ell}(t)$, and ${\boldsymbol{F}}(t)={\boldsymbol{F}}^{\ell}(t)$;\\
		\STATE Update $\mathcal{C}_k(t+1)$ and $\mathcal{C}_{\rm MEC}(t+1)$ according to \eqref{MUqueue} and \eqref{MECqueue};\\
		\ENDFOR
		\ENSURE ${\boldsymbol{w}}$, ${\boldsymbol{x}}$, ${\boldsymbol{F}}$, $\mathcal{C}_k$, and $\mathcal{C}_{\rm MEC}$.
	\end{algorithmic}
\end{algorithm}

\begin{figure}[t]
\centering
\includegraphics[width=0.9\linewidth]{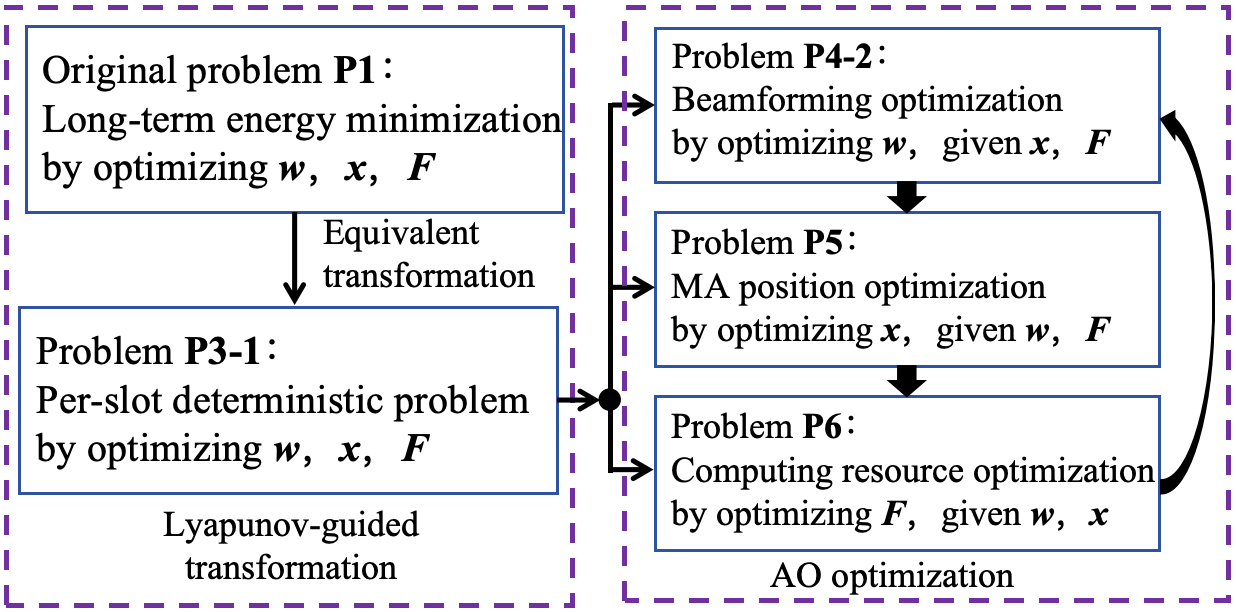}
\caption{{The workflow of the proposed scheme.}}
\label{fig: workflow}
\end{figure}

The overall procedure of the proposed scheme is detailed in Algorithm 1.
{We also present the workflow of the proposed scheme as illustrated in Fig. \ref{fig: workflow}.}

{
\paragraph{Convergence Analysis}
The convergence of the proposed algorithm consists of two aspects, i.e.,
the stability of Lyapunov-guided long-term optimization performance as presented in Theorem \ref{thm:lyapunov}, and the convergence within each time slot via AO as introduced in Theorem \ref{thm:ao_convergence}.

\begin{theorem}
\label{thm:lyapunov}
(\textit{Stability of long-term optimization}):
The long-term buffer stability of MUs and the MEC server and the upper bound of the system energy consumption hold:
\begin{align}
\lim_{T \to \infty} \frac{1}{T} \sum_{t=1}^{T} \mathbb{E}\left\{ \|\mathcal{C}(t)\|_1 \big| \mathcal{C}(t) \right\} 
&\leq \frac{\tilde{\mathcal{C}} + \Gamma E^{*}}{\epsilon},
\end{align}
\begin{align}
\lim_{T \to \infty} \frac{1}{T} \sum_{t=1}^{T} \mathbb{E}\left\{ \sum_{k= 1}^{K}E_{k}(t) + E_{\mathrm{MEC}}(t) \right\} \leq \frac{\tilde{\mathcal{C}}}{\Gamma} + E^{*},\end{align}
where $E^{*} = \sum_{k=1}^{K} E_k^{*} + E_{\text{MEC}}^{*}$ represents the optimal energy consumption achievable by any stationary randomized policy \cite{10592015}.
\end{theorem}

\begin{proof}
Please refer to Appendix \ref{appthmlya}.
\end{proof}

\begin{theorem}
\label{thm:ao_convergence}
(\textit{Convergence of AO algorithm}):
Within each time slot $t$, the proposed AO algorithm converges to a stationary point of problem \textbf{P3-1}.
\end{theorem}

\begin{proof}
Please refer to Appendix \ref{apptmao}.
\end{proof}

\begin{rmk}
The overall convergence is guaranteed by the combination of Theorems \ref{thm:lyapunov} and \ref{thm:ao_convergence}. Lyapunov optimization ensures long-term task buffer stability and bounded energy consumption, while the AO algorithm ensures that each per-slot subproblem is solved efficiently with guaranteed convergence. The balance coefficient $\Gamma$ controls the trade-off between energy consumption and task backlog, as validated in Section V.
\end{rmk}
}

\paragraph{Computational Complexity Analysis}
The computational complexity of Algorithm 1 is mainly determined by the three subproblems solved in each AO iteration, i.e., the receive beamforming optimization problem (\textbf{P4-2}), the MA position optimization problem (\textbf{P5-1}), and the computing resource allocation problem (\textbf{P6}). Let $I_{\rm AO}$ denote the number of AO iterations in each time slot, and $I_{\rm PGA}$ denote the number of PGA iterations required for solving problem (\textbf{P5-1}).

Problem (\textbf{P4-2}) is an SDR problem involving $K$ semi-definite matrix variables of size $M \times M$. Following the standard complexity order of SDP optimization, its computational complexity is $\mathcal{O}(KM^3)$. For problem (\textbf{P5-1}), the PGA method updates the $M$ antenna positions iteratively. In each PGA iteration, the gradient calculation considers the interference coupling among $K$ MUs, resulting in a complexity of $\mathcal{O}(K^2M^2)$. Thus, the total complexity of the PGA procedure is $\mathcal{O}(I_{\rm PGA}K^2M^2)$. Problem (\textbf{P6}) is a convex optimization problem with $K+1$ variables, and its complexity is $\mathcal{O}((K+1)^3)$ when the interior-point method is adopted.
Therefore, the overall computational complexity of Algorithm 1 over $T$ time slots can be expressed as
$\mathcal{O}\left(
T I_{\rm AO}
\left(
KM^3+I_{\rm PGA}K^2M^2+(K+1)^3
\right)
\right)$.

\section{Numerical Results}
In this section, we evaluate the performance of the proposed algorithm through numerical simulations. {The simulation scenario consists of $K = 4$ MUs, each has a maximum local computing capability of $F_k^{\rm{max}} = 3 \times 10^{8}$ Hz, while the maximum computing capability of the MEC server is set to $F^{\rm{max}} = 7 \times 10^{9}$ Hz. The total number of time slots is $T = 500$, and the duration of each time slot is $\tau = 0.5$ s. The input data size of each task $D_k(t)$ is clipped within $[1 \times 10^4, 2 \times 10^5]$ bits, and the required number of computing cycles per bit is set to $\xi = 2000$. The computing energy coefficient is set to $\kappa_{\rm{comp}} = 10^{-26}$, and the balance coefficient $\Gamma$ is set to $3 \times 10^{9}$. The maximum transmit power of each MU is $p_k = 27$ dBm, and the maximum antenna range is $x_{\max} = 12$. The system bandwidth $B=1$ MHz.}

To demonstrate the effectiveness of the proposed algorithm, we compare it with the following benchmark schemes.
\begin{itemize}
    \item \textbf{FPA scheme:} In this scheme, the BS is equipped with conventional FPAs, where the antenna positions remain unchanged during the whole transmission process.
    \item {\textbf{Maximal ratio combining (MRC) beamforming scheme:} This scheme adopts the MRC criterion, where the receive beamforming vector of each user is aligned with the normalized instantaneous channel vector, instead of being obtained by solving the SDR-based beamforming optimization problem.}
    \item {\textbf{Random antenna position (RAP) scheme:} In this scheme, the antenna position is randomly generated.}
\end{itemize}

\begin{figure}[t]
\centering
\includegraphics[width=0.85\linewidth]{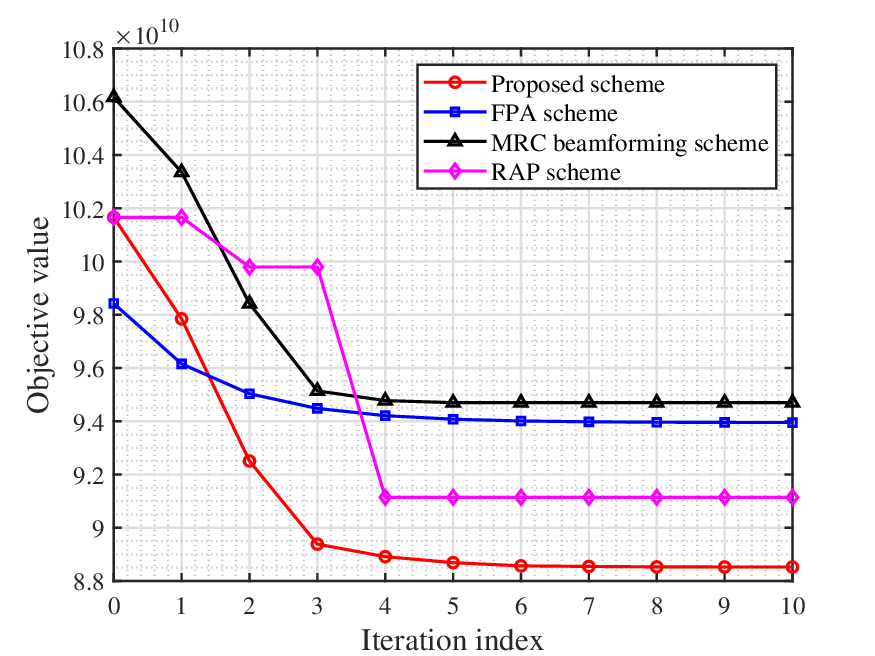}
\caption{{The convergence of all schemes.}}
\label{fig: convergence}
\end{figure}

{Fig. \ref{fig: convergence} illustrates the convergence behavior of the per-slot objective value under different schemes. It can be observed that the objective values of all schemes decrease rapidly and become stable within approximately $6$ iterations, demonstrating the fast numerical convergence of the iterative algorithms. Moreover, the proposed scheme achieves the lowest converged objective value. Since the objective in (25a) jointly incorporates the queue-weighted drift terms and the energy-consumption penalty, a lower objective value indicates a more effective balance between task-buffer reduction and system energy consumption. This performance gain is attributed to the joint optimization of receive beamforming, MA positions, and computing-resource allocation. In contrast, the FPA scheme lacks spatial reconfigurability, the RAP scheme cannot guarantee effective antenna positions, and the MRC beamforming scheme has limited capability in suppressing multi-user interference.}

\begin{figure}[t]
\centering
\includegraphics[width=0.85\linewidth]{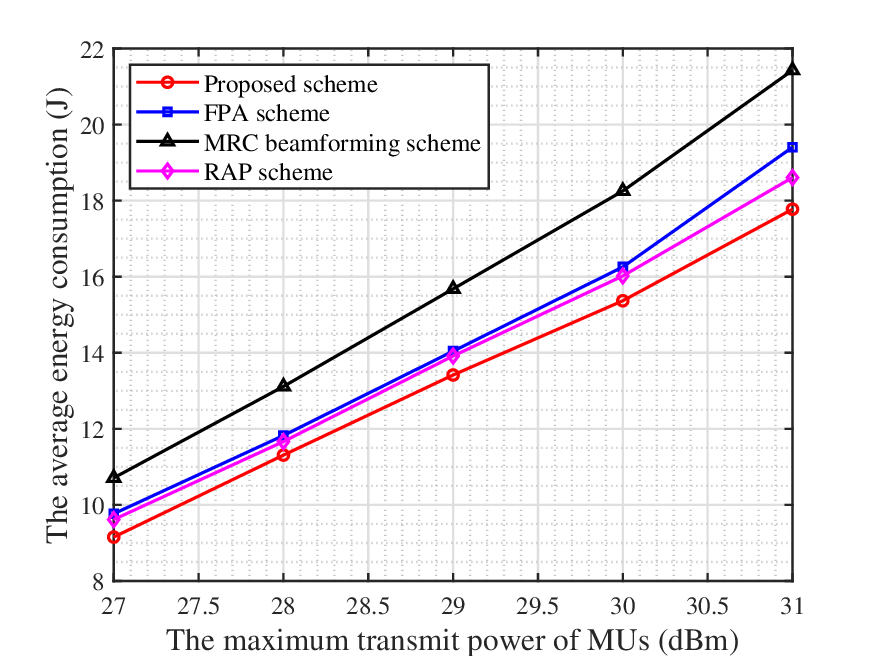}
\caption{{The average energy consumption versus the maximum transmit power of MUs.}}
\label{fig: transmit_power}
\end{figure}

\begin{figure}[t]
\centering
\includegraphics[width=0.85\linewidth]{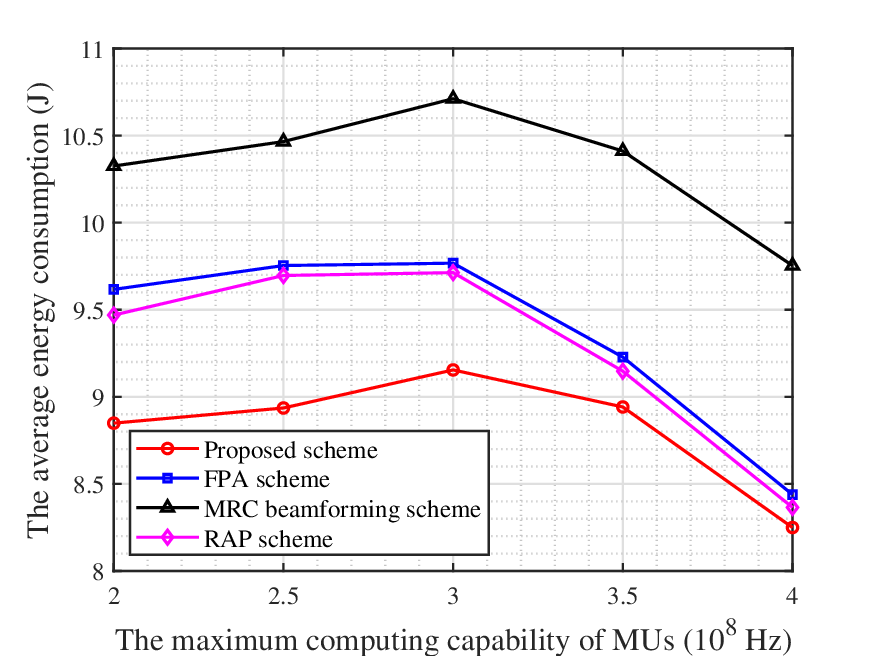}
\caption{{The average energy consumption versus the maximum computing capability of MUs.}}
\label{fig: IoT_CPU}
\end{figure}

\begin{figure}[t]
\centering
\includegraphics[width=0.85\linewidth]{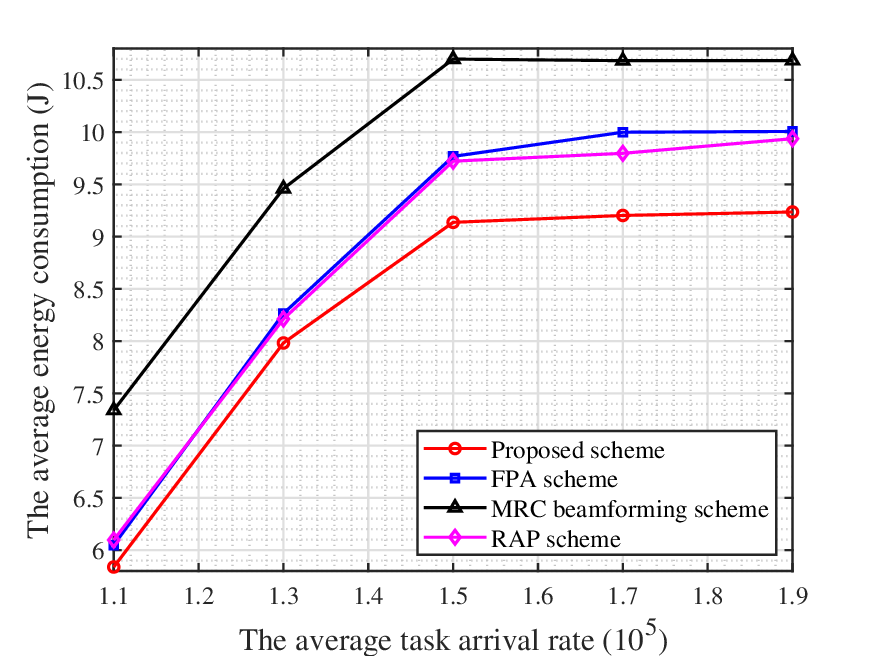}
\caption{{The average energy consumption versus the mean task-arrival data size of MUs.}}
\label{fig: rho_D}
\end{figure}

{Fig. \ref{fig: transmit_power} shows the relationship between the average energy consumption and the maximum transmit power of MUs. It can be observed that the average energy consumption of all schemes increases as the maximum transmit power becomes larger. The main reason is that the offloading energy consumption of each MU is directly proportional to its transmit power, i.e., \(E_{k}^{\text{off}}(t)=p_{k}\tau\). In contrast, the achievable rate only increases logarithmically with the signal-to-interference-plus-noise ratio, and thus the improvement in the offloading rate is relatively limited. Furthermore, when the transmit powers of all MUs increase simultaneously, both the desired signal power and the multi-user interference power are enhanced, which limits the transmission performance improvement brought by increasing transmit power and consequently leads to a higher average system energy consumption.
Moreover, the proposed scheme always achieves the lowest average energy consumption. This is because the proposed scheme jointly optimizes the receive beamforming and MA positions, thereby fully exploiting the spatial DoFs of the wireless channel, improving the effective channel gain, and enhancing multi-user interference suppression. As a result, the task offloading efficiency is improved and unnecessary energy consumption is reduced.}

\begin{figure}[t]
\centering
\includegraphics[width=0.85\linewidth]{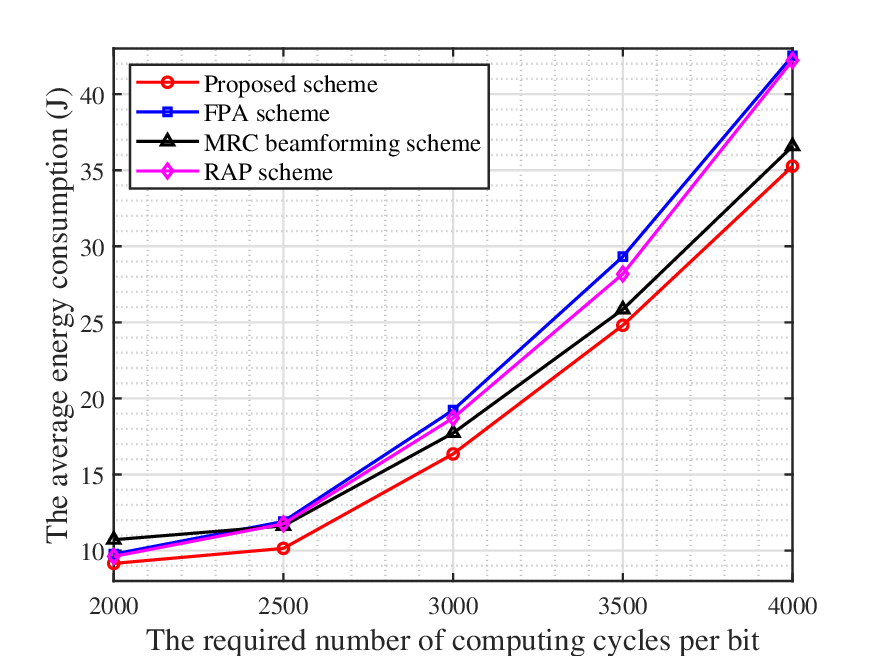}
\caption{{The average energy consumption versus the number of computing cycles required to process a 1-bit task.}}
\label{fig: CRC}
\end{figure}

\begin{figure}[t]
\centering
\includegraphics[width=0.85\linewidth]{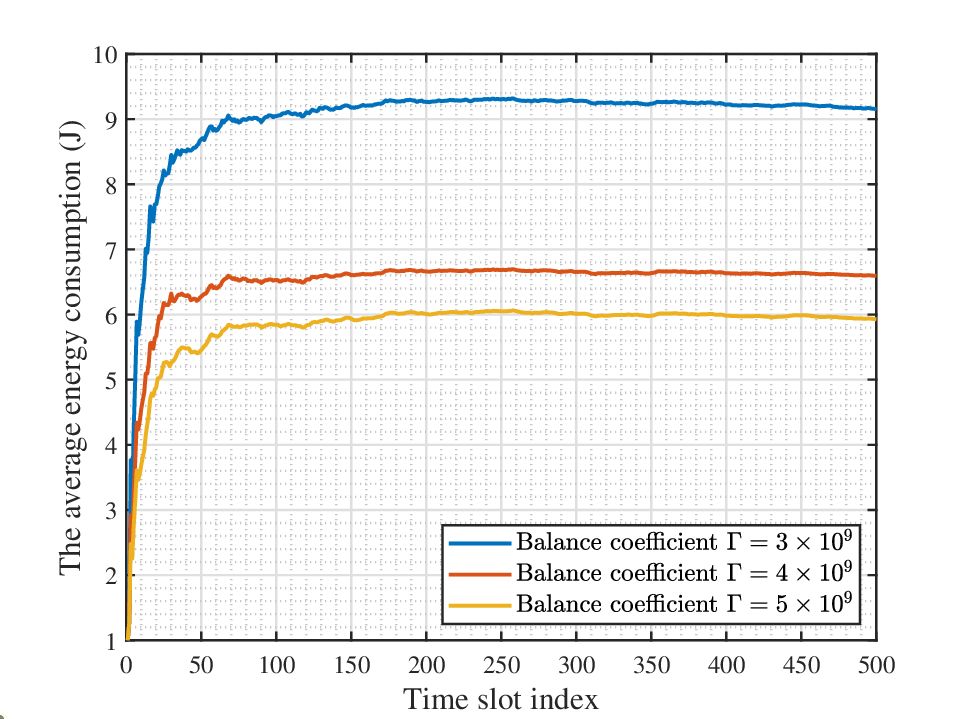}
\caption{{The average energy consumption versus the time slot index.}}
\label{fig: energy_consumption}
\end{figure}

{Fig. \ref{fig: IoT_CPU} illustrates the relationship between the average energy consumption and the maximum computing capability of MUs. As the maximum computing capability of MUs increases, the feasible region for local computing resource allocation is enlarged, allowing the system to process more buffered tasks locally. Therefore, when additional local computing resources are utilized to reduce task backlogs and maintain queue stability, the system energy consumption tends to increase. When the maximum computing capability becomes sufficiently large, the Lyapunov-based resource allocation strategy proposed in this paper can schedule local computing resources more flexibly and avoid excessively high computing frequencies, thereby reducing the average system energy consumption. This phenomenon reflects the performance tradeoff between energy minimization and queue stability in the considered system.}

{Fig. \ref{fig: rho_D} shows the average energy consumption versus the average task-arrival data size of MUs, i.e., \(\rho_\text{D}\). It can be observed that the average energy consumption increases as \(\rho_\text{D}\) becomes larger. This is because a higher task arrival intensity introduces a heavier computational workload into the task buffers of both MUs and the MEC server. Consequently, the system needs to allocate more local computing resources, edge computing resources, and offloading transmission resources to maintain queue stability. When \(\rho_\text{D}\) lies in a relatively large-value region, the growth rate of energy consumption gradually slows down. This is because the computing capabilities and transmission resources are bounded, and most newly arrived tasks accumulate in the task buffers rather than being completely processed immediately. Moreover, under different task arrival intensities, the proposed scheme always achieves the lowest average energy consumption, which verifies that the proposed scheme can efficiently handle bursty task traffic through joint flexible beamforming, MA position optimization, and queue-aware resource allocation.}

\begin{figure}[t]
\centering
\includegraphics[width=0.85\linewidth]{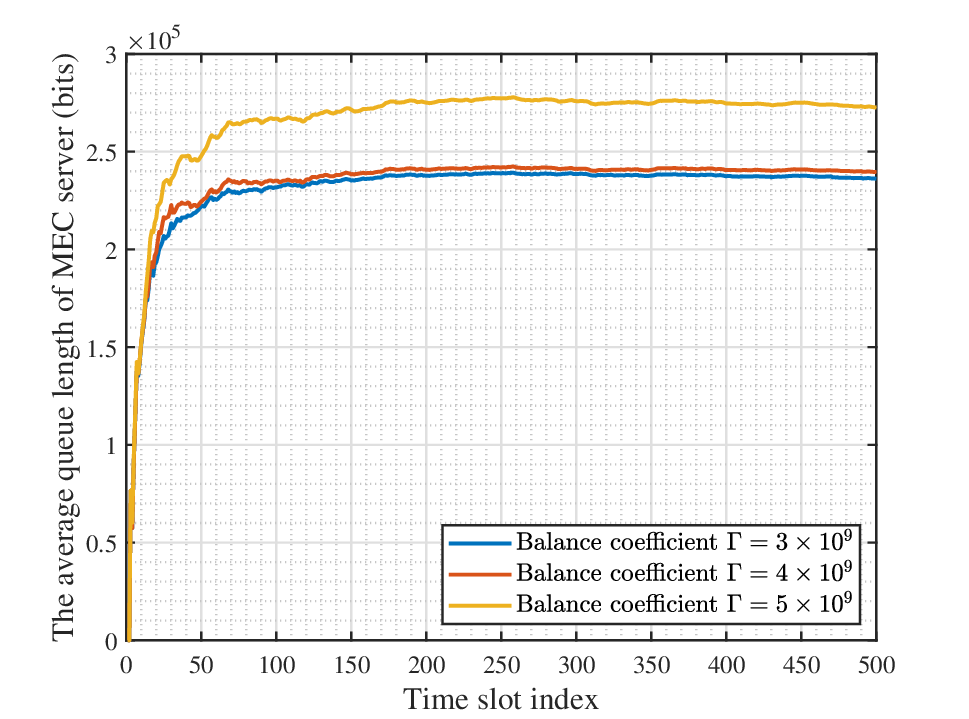}
\caption{{The average queue length of the MEC server versus the time slot index.}}
\label{fig: MEC_len}
\end{figure}

\begin{figure}[t]
\centering
\includegraphics[width=0.85\linewidth]{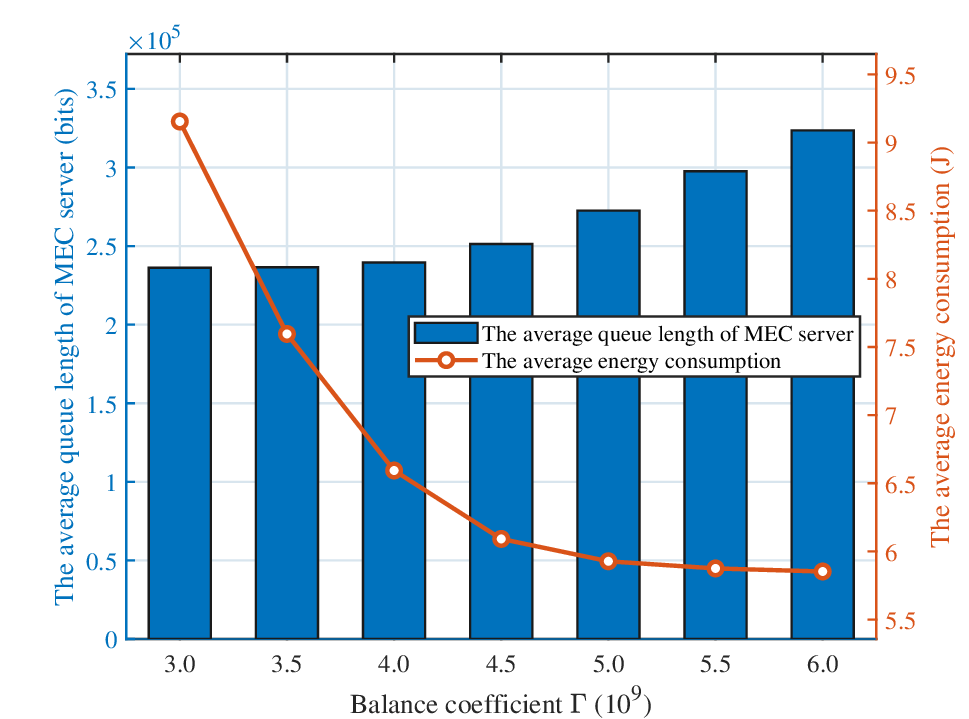}
\caption{{The tradeoff between the average queue length of the MEC server and the average energy consumption.}}
\label{fig: trade_off}
\end{figure}

{Fig. \ref{fig: CRC} presents the average energy consumption versus the number of computing cycles required to process a 1-bit task. It can be observed that, for all schemes, the average energy consumption increases significantly as the required number of computing cycles per bit becomes larger. This is because a higher computing-cycle requirement substantially increases the computation intensity of 1-bit task data. Accordingly, both MUs and the MEC server need to allocate more computing resources to process the same amount of task data, which eventually leads to a rapid increase in computing energy consumption.
Moreover, the proposed scheme consistently outperforms all benchmark schemes in terms of energy consumption. This indicates that the joint optimization of receive beamforming, MA positions, and computing resource allocation can effectively reduce the system energy consumption even when the task computation intensity is high.}

Fig. \ref{fig: energy_consumption} depicts the average energy consumption versus the time slot index under different balance coefficients $\Gamma$. It can be observed that, although the instantaneous energy consumption fluctuates over time due to stochastic task arrivals and the time-varying wireless environment, the average energy consumption remains within a relatively stable range for each considered value of $\Gamma$. This indicates that the proposed online optimization framework can maintain stable long-term system operation under dynamic environments. Moreover, the average energy consumption decreases as the balance coefficient increases from $\Gamma = 3 \times 10^{9}$ to $\Gamma = 5 \times 10^{9}$. The reason is that a larger balance coefficient $\Gamma$ assigns a higher weight to the energy minimization term in the drift-plus-penalty objective, thereby driving the system to adopt more energy-efficient decisions in receive beamforming optimization, MA position adjustment, and computing resource allocation. As a result, the system operates with a lower long-term energy expenditure. Therefore, Fig. \ref{fig: energy_consumption} clearly demonstrates that the proposed scheme is capable of effectively regulating the energy consumption behavior of the system through the balance coefficient $\Gamma$, while preserving stable operation over a long time horizon.

Fig. \ref{fig: MEC_len} shows the average queue length of the MEC server versus the time slot index under different balance coefficients $\Gamma$. It can be observed that the MEC queue length fluctuates over time for all considered values of $\Gamma$, due to the randomness of task arrivals and the time-varying wireless environment. Nevertheless, the queue length remains bounded within a stable range throughout the entire time horizon, which confirms that the proposed scheme is capable of maintaining queue stability under dynamic system conditions. Moreover, the average queue length of the MEC server increases as the balance coefficient rises from $\Gamma = 3 \times 10^{9}$ to $\Gamma = 5 \times 10^{9}$. This is because a larger balance coefficient places more emphasis on energy minimization in the drift-plus-penalty objective, leading the system to adopt more conservative resource allocation and task processing decisions, and thus allowing more offloaded task data to accumulate in the MEC queue.

Fig. \ref{fig: trade_off} illustrates the tradeoff between the average queue length of the MEC server and the average energy consumption under different balance coefficients $\Gamma$. It is observed that, as $\Gamma$ increases, the average energy consumption decreases monotonically, whereas the average queue length of the MEC server increases gradually. This trend is expected because a larger balance coefficient assigns a higher priority to energy minimization in the drift-plus-penalty objective, which drives the system to adopt more energy-efficient transmission, antenna adjustment, and computing resource allocation. However, such conservative decisions also reduce the aggressiveness of task processing, thereby causing more task data to accumulate in the MEC queue. Therefore, Fig. \ref{fig: trade_off} clearly reveals the fundamental energy-queue tradeoff achieved by the proposed scheme. In particular, by tuning the balance coefficient, the system can flexibly operate at different points on the tradeoff curve according to practical design requirements.

\section{Conclusion}
This paper investigates an energy-efficient MEC system enhanced by MAs and task buffering. Addressing the limitations of FPAs and rigid single-slot execution, we propose a framework that jointly optimizes the receive beamforming, the MA positions, and the computing resource allocation to minimize the long-term energy consumption.
By integrating task buffer queues at both MUs and the BS, our proposed scheme allows tasks to be stored upon arrival and scheduled flexibly across multiple time slots for bursty workloads typical of generative AI and LAE services.
We formulate a stochastic optimization problem subject to the task buffer queue stability constraints and employ Lyapunov optimization to decouple it into deterministic per-slot subproblems. An efficient AO algorithm is then developed, utilizing the SDR for the beamforming optimization and the PGA for the MA positioning.
Extensive simulations validate the proposed scheme, demonstrating lower energy consumption than the benchmark schemes. The results also confirm that the proposed framework maintains stable task buffer queues and effectively balances energy consumption and queue backlog through adjustable control parameters.

\begin{appendices}

\section{Proof of Proposition \ref{propositon2}}
\label{appprp1}
    \begin{proof}
        For any two adjacent antennas, we have
        $x_{\min} \leq x_m(t) - x_{m-1}(t)$ to avoid the coupling effect between the antennas, which is equivalent to
        \begin{equation}
        \begin{split}
            x_1(t) &+ (M-1)x_{\min} \\ 
            &\leq x_2(t) + (M-2)x_{\min}  \\ 
            &\leq \cdots \leq x_m(t) + (M-m)x_{\min} \\ 
            & \leq \cdots \leq x_{M-2}(t) + 2x_{\min} \\ 
            & \leq x_{M-1}(t) + x_{\min} \leq x_M(t).
            \label{p22}
        \end{split}
        \end{equation}
        Since $x_M(t) \leq x_{\max}$, we have
        \begin{equation}
            x_m(t) + (M-m)x_{\min} \leq x_{\max}.
            \label{p23}
        \end{equation}
        After simple transformations of \eqref{p22} and \eqref{p23}, we can thus obtain \eqref{p21}. For simplicity, we define $x_0  = -x_{\min}$.
    \end{proof}

\section{Proof of Lemma \ref{le1}}
\label{apple1}
\begin{proof}
Since $(\max\{a-b,0\}+c)^2 \leq a^2+b^2+c^2+2a(c-b)$, while $a,b,c\geq 0$, we have
\begin{equation}
\begin{split}
    (\mathcal{C}_{\mathrm{MEC}}(t+1))^2 
    \leq&\
    (\mathcal{C}_{\mathrm{MEC}}(t))^2 + (D_{\mathrm{MEC}}(t))^2 + ({D}_{\mathrm{MEC}}^{\mathrm{off}}(t))^2\\
    &+ 2\mathcal{C}_{\mathrm{MEC}}(t) (D_{\mathrm{MEC}}^{\mathrm{off}}(t)-D_{\mathrm{MEC}}(t)).
\end{split}
\label{le11}
\end{equation}
Similarly, we have
\begin{equation}
\begin{split}
    (\mathcal{C}_{k}(t+1))^2 \leq&\
    (\mathcal{C}_{k}(t))^2 + (D_k^{\mathrm{out}}(t))^2 + ({D}_{k}^{\mathrm{}}(t))^2 \\&+
    2\mathcal{C}_{k}(t)(D_k(t)-D_k^{\mathrm{out}}(t)).
\end{split}
\label{le12}
\end{equation}
Since the data sizes of arrived tasks are bounded, and the communication and computing resources are limited, there always exist $D^{\max} \geq D_k (t)$, $D^{\mathrm{out}, \max} \geq D_k^{\mathrm{out}} (t)$, $D_{\mathrm{off}, \mathrm{MEC}}^{\max} \geq D_{\mathrm{MEC}}^{\mathrm{off}}(t)$, and $D_{\mathrm{MEC}}^{\max} \geq D_{\mathrm{MEC}} (t)$. Hence, we have
\begin{equation}
\begin{split}
    &(D_{\mathrm{MEC}}^{\mathrm{off}}(t))^2+(D_{\mathrm{MEC}}^{\mathrm{}}(t))^2 + 
    \sum_{k=1}^K((D_k (t))^2+(D_k^{\mathrm{out}} (t))^2)
    \\ &\leq (D_{\mathrm{off}, \mathrm{MEC}}^{\max})^2 + (D_{\mathrm{MEC}}^{\max})^2 + K((D^{\max})^2+(D^{\mathrm{out}, \max})^2)\\
    & \triangleq 2\Tilde{\mathcal{C}}.
\end{split}
\label{le13}
\end{equation}
By substituting \eqref{le11}, \eqref{le12}, and 
\eqref{le13} into \eqref{drift}, we can obtain \eqref{driftlower}.
\end{proof}

\section{Proof of Proposition \ref{pro2pga}}
\label{appgpa}
\begin{proof}
    Firstly, the gradient $\nabla_{\boldsymbol{x}^{j}(t)}
    \Tilde{\Psi}\left( \boldsymbol{a}_i(t), \boldsymbol{b}_i(t) \vert_{i\in \Tilde{\mathcal{K}}} \right )$ with respect to $\boldsymbol{x}(t)$ is given by \eqref{gd1}, which is presented at the top of the next page, and the time slot notations in \eqref{gd1} and the rest of this proof are omitted, for simplicity, where the gradient $\nabla_{\boldsymbol{x}^{j}(t)}  \Psi_i (\boldsymbol{a}_i(t), \boldsymbol{b}_i(t);\boldsymbol{w}_k(t) )$ is given by
    \begin{figure*}[!t]
    \vspace*{-\baselineskip} 
    \begin{equation}{\small
    \begin{split}
    \nabla_{\boldsymbol{x}^{j}(t)}
    \Tilde{\Psi}\left( \boldsymbol{a}_i(t), \boldsymbol{b}_i(t) \vert_{i\in \Tilde{\mathcal{K}}} \right )
    =
    \sum_{k\in \Tilde{\mathcal{K}}} \Tilde{\Upsilon}_k \left(   \frac{\sum_{i\in \Tilde{\mathcal{K}}} \nabla_{\boldsymbol{x}^{j}}  \Psi_i (\boldsymbol{a}_i, \boldsymbol{b}_i;\boldsymbol{w}_k )}{\ln(2)\left( \sum_{i\in \Tilde{\mathcal{K}}}  \Psi_i (\boldsymbol{a}_i, \boldsymbol{b}_i;\boldsymbol{w}_k ) + \Tilde{w}_k \right)}-  \frac{ \sum_{i\in \Tilde{\mathcal{K}}, i \neq k} \nabla_{\boldsymbol{x}^{j}}  \Psi_i (\boldsymbol{a}_i, \boldsymbol{b}_i;\boldsymbol{w}_k ) }{\ln(2) \left( \sum_{i\in \Tilde{\mathcal{K}}, i \neq k}  \Psi_i (\boldsymbol{a}_i, \boldsymbol{b}_i;\boldsymbol{w}_k ) + \Tilde{w}_k \right ) }
 \right),
\end{split}
\label{gd1}}
    \end{equation}
    \hrulefill
    \end{figure*}
\begin{equation}
\begin{split}
    &\nabla_{\boldsymbol{x}^{j}}  \Psi_i (\boldsymbol{a}_i, \boldsymbol{b}_i;\boldsymbol{w}_k ) \\&=
    \left[
    \frac{\partial \Psi_i (\boldsymbol{a}_i, \boldsymbol{b}_i;\boldsymbol{w}_k ) }{\partial x_1^j }, \cdots,
    \frac{\partial \Psi_i (\boldsymbol{a}_i, \boldsymbol{b}_i;\boldsymbol{w}_k ) }{\partial x_M^j }
    \right]^{\mathsf{T}},
\end{split}
\label{gd2}
\end{equation}
For any $m \in \mathcal{M}$, $\frac{\partial \Psi_i (\boldsymbol{a}_i, \boldsymbol{b}_i;\boldsymbol{w}_k ) }{\partial x_m^j }$ can be expressed as
\begin{equation}
\begin{split}
    &\frac{\partial \Psi_i (\boldsymbol{a}_i, \boldsymbol{b}_i;\boldsymbol{w}_k ) }{\partial x_m^j }  = \ 
    \frac{\partial \Psi_i (\boldsymbol{a}_i, \boldsymbol{b}_i;\boldsymbol{w}_k ) }{\partial (a_i^m)^j } \frac{\partial (a_i^m)^j}{\partial x_m^j} \\ &\quad\quad\quad\quad\quad\quad\quad\quad + 
    \frac{\partial \Psi_i (\boldsymbol{a}_i, \boldsymbol{b}_i;\boldsymbol{w}_k ) }{\partial (b_i^m)^j } \frac{\partial (b_i^m)^j}{\partial x_m^j} \\ & = 
    \frac{-2\pi}{\lambda} \cos\Phi_{i \rightarrow \mathrm{BS}} \sin \left (
        \frac{2\pi x_m}{\lambda}\cos \Phi_{i \rightarrow \mathrm{BS}} 
    \right ) \frac{\partial \Psi_i (\boldsymbol{a}_i, \boldsymbol{b}_i;\boldsymbol{w}_k ) }{\partial (a_i^m)^j } \\ & +
    \frac{2\pi}{\lambda} \cos\Phi_{i \rightarrow \mathrm{BS}} \cos \left (
        \frac{2\pi x_m}{\lambda}\cos \Phi_{i \rightarrow \mathrm{BS}} 
    \right ) \frac{\partial \Psi_i (\boldsymbol{a}_i, \boldsymbol{b}_i;\boldsymbol{w}_k ) }{\partial (b_i^m)^j }.
\end{split}
\label{partial1}
\end{equation}
By substituting \eqref{partial1} into \eqref{gd2}, we can derive
\begin{equation}
\begin{split}
    &\nabla_{\boldsymbol{x}^{j}}  \Psi_i (\boldsymbol{a}_i, \boldsymbol{b}_i;\boldsymbol{w}_k ) \\&=
    \frac{1}{2}\boldsymbol{A}_i \nabla_{\boldsymbol{a}_i^j} \Psi_i (\boldsymbol{a}_i, \boldsymbol{b}_i;\boldsymbol{w}_k )
    +\frac{1}{2} \boldsymbol{B}_i \nabla_{\boldsymbol{b}_i^j} \Psi_i (\boldsymbol{a}_i, \boldsymbol{b}_i;\boldsymbol{w}_k )\\
    &=\varphi_i\boldsymbol{A}_i \left(\boldsymbol{U}_k^{(1)}
    \boldsymbol{a}_i + \boldsymbol{U}_k^{(2)} \boldsymbol{b}_i
    \right) +
    \varphi_i\boldsymbol{B}_i \left(\boldsymbol{U}_k^{(1)}
    \boldsymbol{b}_i - \boldsymbol{U}_k^{(2)} \boldsymbol{a}_i
    \right),\\
    &\triangleq \Tilde{\boldsymbol{X}}_i (\boldsymbol{a}_i, \boldsymbol{b}_i;\boldsymbol{w}_k ),
\end{split}
\label{gd3}
\end{equation}
where $\boldsymbol{A}_i$ and $\boldsymbol{B}_i$ are given by \eqref{biga} and \eqref{bigb}, respectively, and given at top of the next page.
By combining \eqref{gd3} and \eqref{gd1}, \eqref{pro1eq1} can be derived.
\begin{figure*}[!t]
\begin{equation}
\boldsymbol{A}_i = \mathsf{diag} \left(
\frac{-4\pi}{\lambda} \cos\Phi_{i \rightarrow \mathrm{BS}} \sin \left (
        \frac{2\pi x_1}{\lambda}\cos \Phi_{i \rightarrow \mathrm{BS}} 
    \right ), \cdots,
\frac{-4\pi}{\lambda} \cos\Phi_{i \rightarrow \mathrm{BS}} \sin \left (
        \frac{2\pi x_M}{\lambda}\cos \Phi_{i \rightarrow \mathrm{BS}} 
    \right )
\right ),
\label{biga}
\end{equation}
\end{figure*}
\begin{figure*}[!t]
\begin{equation}
\boldsymbol{B}_i = \mathsf{diag} \left(
\frac{4\pi}{\lambda} \cos\Phi_{i \rightarrow \mathrm{BS}} \cos \left (
        \frac{2\pi x_1}{\lambda}\cos \Phi_{i \rightarrow \mathrm{BS}} 
    \right ), \cdots,
\frac{4\pi}{\lambda} \cos\Phi_{i \rightarrow \mathrm{BS}} \cos \left (
        \frac{2\pi x_M}{\lambda}\cos \Phi_{i \rightarrow \mathrm{BS}} 
    \right )
\right ),
\label{bigb}
\end{equation}
\hrulefill
\end{figure*}
\end{proof}

{
\section{Proof of Theorem \ref{thm:lyapunov}}
\label{appthmlya}
\begin{proof}
    Firstly, following inequality (64) in \cite{10592015}, the upper bound of the drift-plus-penalty $\Phi(\mathcal{C}(t))$ can be written as
\begin{align}
\Phi(\mathcal{C}(t)) &\leq \mathcal{C}_{\mathrm{MEC}} (t) \mathbb{E}[D_{\mathrm{MEC}}^{\mathrm{off}}(t)-D_{\mathrm{MEC}}(t)\vert \mathcal{C}(t)]\nonumber\\
    &\quad + \sum_{k=1}^K \mathcal{C}_{k} (t) \mathbb{E}[D_{k}^{\mathrm{}}(t)-D_{k}^{\mathrm{out}}(t)\vert \mathcal{C}(t)] \nonumber\\&\quad +  \tilde{\mathcal{C}} +
    \Gamma \mathbb{E}\left[\sum_{k= 1}^{K}E_{k}(t) + E_{\mathrm{MEC}}(t)\vert \mathcal{C}(t)\right] \nonumber\\
&\leq \tilde{\mathcal{C}} - \epsilon \cdot \mathbb{E}\left[ {\mathcal{C}}_{\text{MEC}}(t) + \sum_{k=1}^{K} {\mathcal{C}}_k(t) \bigg| \mathcal{C}(t) \right] \nonumber \\
&\quad + \Gamma \left(\sum_{k=1}^{K} E_k^{*} + E_{\text{MEC}}^{*}\right) \nonumber \\
&= \tilde{\mathcal{C}} - \epsilon \cdot \mathbb{E}\left\{ \|\mathcal{C}(t)\|_1 \big| \mathcal{C}(t) \right\} + \Gamma E^{*}, \label{eq:app5}
\end{align}
where $\epsilon$ is a positive constant, $\|\mathcal{C}(t)\|_1 = {\mathcal{C}}_{\text{MEC}}(t) + \sum_{k=1}^{K} {\mathcal{C}}_k(t)$, and $E^{*} = \sum_{k=1}^{K} E_k^{*} + E_{\text{MEC}}^{*}$ represents the optimal energy consumption achievable by any stationary randomized policy \cite{10592015}. For any time slot $t \in \mathcal{T}' \triangleq \{t | t \neq T, t \in \mathcal{T}\}$, we can obtain
\begin{align}
&\mathbb{E}\{ \mathcal{L}(\mathcal{C}(T)) -  \mathcal{L}(\mathcal{C}(0))\} + \sum_{t \in \mathcal{T}'} \mathbb{E}\left\{ \|\mathcal{C}(t)\|_1 \big| \mathcal{C}(t) \right\} \nonumber \\
&\leq T \tilde{\mathcal{C}} - \sum_{t \in \mathcal{T}'} \mathbb{E}\left\{ \|\mathcal{C}(t)\|_1 \big| \mathcal{C}(t) \right\} + T \Gamma E^{*}. \label{eq:app6}
\end{align}
Since the initial data size of the task buffer is empty, i.e., $\mathbb{E}\{\mathcal{L}(\mathcal{C}(0)) \} = 0$, and the buffer will be stable at time slot $T$, i.e., $\mathbb{E}\{\mathcal{L}(\mathcal{C}(T)) \}$ is the bounded constants.
Dividing both sides of \eqref{eq:app6} by $T\epsilon$ and then letting $T \to \infty$, we obtain that
\begin{align}
\lim_{T \to \infty} \frac{1}{T} \sum_{t=1}^{T} \mathbb{E}\left\{ \|\mathcal{C}(t)\|_1 \big| \mathcal{C}(t) \right\} 
&\leq \frac{\tilde{\mathcal{C}} + \Gamma E^{*}}{\epsilon} . \label{eq:app7}
\end{align}

Similarly, dividing both sides of \eqref{eq:app6} by $T\Gamma$ and then letting $T \to \infty$, we can obtain
\begin{align}
\lim_{T \to \infty} \frac{1}{T} \sum_{t=1}^{T} \mathbb{E}\left\{ \sum_{k= 1}^{K}E_{k}(t) + E_{\mathrm{MEC}}(t) \right\} \leq \frac{\tilde{\mathcal{C}}}{\Gamma} + E^{*}. \label{eq:app8}
\end{align}
Ineqs. \eqref{eq:app7} and \eqref{eq:app8} indicates that the task buffers of MUs and the MEC server are stable, and the system energy consumption is upper bounded.
Then we complete the proof of Theorem \ref{thm:lyapunov}.
\hfill $\square$
\end{proof}

\section{Proof of Theorem \ref{thm:ao_convergence}}
\label{apptmao}
\begin{proof}
    Let $\Phi \left(\boldsymbol{w}^\ell(t), \boldsymbol{x}^\ell(t), \boldsymbol{F}^\ell(t) \right)$ denote the objective value of problem \textbf{P3-1} at the $\ell$-th AO iteration in time slot $t$. Accordingly, in each iteration of Algorithm \ref{algoverall}, we first optimize $\boldsymbol{w}^\ell(t)$ with fixed $\boldsymbol{x}^\ell(t)$ and $\boldsymbol{F}^\ell(t)$ by solving problem \textbf{P4-2} via Step 5 to obtain $\boldsymbol{w}^{\ell+1}(t)$. Hence, we have
\begin{equation}
    \Phi \left(\boldsymbol{w}^{\ell+1}(t), \boldsymbol{x}^\ell(t), \boldsymbol{F}^\ell(t) \right)
    \leq
    \Phi \left(\boldsymbol{w}^\ell(t), \boldsymbol{x}^\ell(t), \boldsymbol{F}^\ell(t) \right).
\end{equation}
Then, we optimize $\boldsymbol{x}^\ell(t)$ with fixed $\boldsymbol{w}^{\ell+1}(t)$ and $\boldsymbol{F}^\ell(t)$ using PGA method via Step 6 to obtain $\boldsymbol{x}^{\ell+1}(t)$. We have
\begin{equation}
    \Phi \left(\boldsymbol{w}^{\ell+1}(t), \boldsymbol{x}^{\ell+1}(t), \boldsymbol{F}^\ell(t) \right)
    \leq
    \Phi \left(\boldsymbol{w}^{\ell+1}(t), \boldsymbol{x}^\ell(t), \boldsymbol{F}^\ell(t) \right).
\end{equation}
Lastly, we optimize $\boldsymbol{F}^\ell(t)$ with fixed $\boldsymbol{w}^{\ell+1}(t)$ and $\boldsymbol{x}^{\ell+1}(t)$ by solving problem \textbf{P6} via Step 7 to obtain $\boldsymbol{F}^{\ell+1}(t)$. We have
\begin{equation}
\begin{split}
    &\Phi \left(\boldsymbol{w}^{\ell+1}(t), \boldsymbol{x}^{\ell+1}(t), \boldsymbol{F}^{\ell+1}(t) \right) \\
    &\leq
    \Phi \left(\boldsymbol{w}^{\ell+1}(t), \boldsymbol{x}^{\ell+1}(t), \boldsymbol{F}^\ell(t) \right).
\end{split}
\end{equation}
Therefore, we can observe that the objective value is non-increasing across iterations.

Meanwhile, the objective function \eqref{objp31} consists of the task buffer drift $\mathcal{C}_{\mathrm{MEC}} (t) \left(D_{\mathrm{MEC}}^{\mathrm{off}}(t)-D_{\mathrm{MEC}}(t)\right) + \sum_{k=1}^K \mathcal{C}_{k} (t) \left(D_{k}^{\mathrm{}}(t)-D_{k}^{\mathrm{out}}(t)\right)$, which is bounded below since all data sizes are non-negative and bounded,
and the energy consumption $\Gamma\left(\sum_{k=1}^{K} E_k(t) + E_{\mathrm{MEC}}(t)\right) \geq 0$, which is non-negative. Thus, $\Phi(\mathcal{C}(t)) \geq 0$, i.e., the objective is lower bounded. By the monotone convergence theorem, a non-increasing sequence that is lower bounded must converge. Therefore, $\Phi^\ell(\mathcal{C}(t))$ converges to a finite value as $\ell \to \infty$.
Hence, the AO algorithm is capable of converging to a stationary point of problem \textbf{P3-1}. This completes the proof of Theorem \ref{thm:ao_convergence}.
\end{proof}
}

\end{appendices}

\bibliographystyle{IEEEtran}
\bibliography{ref}

@ARTICLE{9252924,
  author={Ren, Jinke and Yu, Guanding and Ding, Guangyao},
  journal={IEEE J. Sel. Areas Commun.}, 
  title={Accelerating {DNN} Training in Wireless Federated Edge Learning Systems}, 
  year={Jan. 2021},
  volume={39},
  number={1},
  pages={219-232},
}

@ARTICLE{10422876,
  author={Cao, Xiaowen and Lyu, Zhonghao and Zhu, Guangxu and Xu, Jie and Xu, Lexi and Cui, Shuguang},
  journal={IEEE Wireless Commun.}, 
  title={An Overview on Over-the-Air Federated Edge Learning}, 
  year={Jun. 2024},
  volume={31},
  number={3},
  pages={202-210},
}

@ARTICLE{11106440,
  author={Du, Hongyang and Niyato, Dusit and Kang, Jiawen and Xiong, Zehui and Lam, Kwok-Yan and Fang, Yuguang and Li, Yonghui},
  journal={IEEE Netw.}, 
  title={Spear or Shield: The Role of Generative {AI} in Intelligent Network Services}, 
  year={Jan. 2026},
  volume={40},
  number={1},
  pages={312-321},
}

@ARTICLE{wu2025low,
  author={Wu, Jun and Yang, Yaoqi and Yuan, Weijie and Liu, Wenchao and Wang, Jiacheng and Mao, Tianqi and Zhou, Lin and Cui, Yuanhao and Liu, Fan and Sun, Geng and others},
  journal={China Commun.}, 
  title={Low-altitude wireless networks: A comprehensive survey}, 
  year={Mar. 2026},
  volume={23},
  number={3},
  pages={99-141},
}

@ARTICLE{10734319,
  author={Wu, Yinyu and Zhang, Xuhui and Ren, Jinke and Xing, Huijun and Shen, Yanyan and Cui, Shuguang},
  journal={IEEE Netw. Lett.}, 
  title={Latency-Aware Resource Allocation for Mobile Edge Generation and Computing via Deep Reinforcement Learning}, 
  year={Dec. 2024},
  volume={6},
  number={4},
  pages={237-241},
}

@ARTICLE{9214878,
  author={Chen, Ying and Liu, Zhiyong and Zhang, Yongchao and Wu, Yuan and Chen, Xin and Zhao, Lian},
  journal={IEEE Trans. Ind. Informat.}, 
  title={Deep Reinforcement Learning-Based Dynamic Resource Management for Mobile Edge Computing in Industrial Internet of Things}, 
  year={Jul. 2021},
  volume={17},
  number={7},
  pages={4925-4934},
}

@INPROCEEDINGS{10012694,
  author={Zhang, Xuhui and Xing, Huijun and Zang, Weilin and Jin, Zhenzhen and Shen, Yanyan},
  booktitle={Proc. IEEE Veh. Technol. Conf.}, 
  title={Cybertwin-Driven Multi-Intelligent Reflecting Surfaces aided Vehicular Edge Computing Leveraged by Deep Reinforcement Learning}, 
  year={London, UK, Sep. 2022},
  volume={},
  number={},
  pages={1-7}
}

@ARTICLE{11154927,
  author={Wang, Dong and Mei, Weidong and Ning, Boyu and Chen, Zhi and Zhang, Rui},
  journal={IEEE Trans. Wireless Commun.}, 
  title={Movable Antenna Enhanced Wide-Beam Coverage: Joint Antenna Position and Beamforming Optimization}, 
  year={2026},
  volume={25},
  number={},
  pages={3541-3558},
}

@ARTICLE{11007274,
  author={Ning, Boyu and Yang, Songjie and Wu, Yafei and Wang, Peilan and Mei, Weidong and Yuen, Chau and Björnson, Emil},
  journal={IEEE Wireless Commun.}, 
  title={Movable Antenna-Enhanced Wireless Communications: General Architectures and Implementation Methods}, 
  year={May 2025},
  volume={32},
  number={5},
  pages={108-116},
}

@ARTICLE{liu2025movable,
  author={Liu, Wenchao and Zhang, Xuhui and Wang, Chunjie and Ren, Jinke and Yuan, Weijie and You, Changsheng},
  journal={IEEE Internet Things Mag.}, 
  title={Fluid Antenna Systems Meet Low-Altitude Wireless Networks: Fundamentals, Opportunities, and Future Directions}, 
  year={May 2026},
  volume={9},
  number={3},
  pages={55-62},
}

@ARTICLE{8766143,
  author={Zhang, Zhengquan and Xiao, Yue and Ma, Zheng and Xiao, Ming and Ding, Zhiguo and Lei, Xianfu and Karagiannidis, George K. and Fan, Pingzhi},
  journal={IEEE Veh. Technol. Mag.}, 
  title={{6G} Wireless Networks: Vision, Requirements, Architecture, and Key Technologies}, 
  year={Sep. 2019},
  volume={14},
  number={3},
  pages={28-41},
}

@INPROCEEDINGS{10693833,
  author={Kuang, Ziming and Liu, Wenchao and Wang, Chunjie and Jin, Zhenzhen and Ren, Jinke and Zhang, Xuhui and Shen, Yanyan},
  booktitle={Proc. IEEE/CIC Int. Conf. Commun. China Workshops (ICCC Workshops)}, 
  title={Movable-Antenna Array Empowered {ISAC} Systems for Low-Altitude Economy}, 
  year={Hangzhou, China, Aug. 2024},
  volume={},
  number={},
  pages={776-781},
}

@ARTICLE{11177504,
  author={Cao, Xiaowen and Jiang, Peng and Zhu, Guangxu and He, Yejun and Guizani, Mohsen},
  journal={IEEE Trans. Netw. Sci. Eng.}, 
  title={Joint Antenna Position and Beamforming Optimization for Movable Antenna-Enabled Secure {IRS}-{ISAC} Network}, 
  year={2026},
  volume={13},
  number={},
  pages={4440-4454},
}

@ARTICLE{11346858,
	author={Zhang, Zeyuan and Xiu, Yue and Dong, Zheng and Yin, Jiacheng and Khabbaz, Maurice J. and Assi, Chadi and Wei, Ning},
	journal={IEEE Wireless Commun. Lett.}, 
	title={Crosstalk-Resilient Beamforming for Movable Antenna Enabled Integrated Sensing and Communication}, 
	year={2026},
	volume={15},
	number={},
	pages={1395-1399},
}

@INPROCEEDINGS{11374087,
  author={Wang, Yike and Wu, Zhike and Chen, Jiang and Wang, Chunjie and Zhang, Xuhui and Shen, Yanyan},
  booktitle={Proc. IEEE Int. Conf. Commun. Technol. (ICCT)}, 
  title={{QoS}-Aware Integrated Sensing, Communication, and Control with Movable Antenna}, 
  year={Shenyang, China, Oct. 2025},
  volume={},
  number={},
  pages={565-570},
}

@INPROCEEDINGS{11374005,
  author={Chen, Yiyang and Liu, Wenchao and Wang, Chunjie and Wu, Yinyu and Zhang, Xuhui and Shen, Yanyan},
  booktitle={Proc. IEEE Int. Conf. Commun. Technol. (ICCT)}, 
  title={Latency Minimization for Multi-{AAV}-Enabled {ISCC} Systems with Movable Antenna}, 
  year={Shenyang, China, Oct. 2025},
  volume={},
  number={},
  pages={1232-1238},
}

@ARTICLE{9968198,
  author={Yu, Jiadong and Li, Yang and Liu, Xiaolan and Sun, Bo and Wu, Yuan and Hin-Kwok Tsang, Danny},
  journal={IEEE Trans. Wireless Commun.}, 
  title={{IRS} Assisted {NOMA} Aided Mobile Edge Computing With Queue Stability: Heterogeneous Multi-Agent Reinforcement Learning}, 
  year={Jul. 2023},
  volume={22},
  number={7},
  pages={4296-4312},
}

@ARTICLE{11328802,
  author={Zhang, Xuhui and Liu, Wenchao and Wang, Chunjie and Ren, Jinke and Xing, Huijun and Wang, Shuqiang and Shen, Yanyan},
  journal={IEEE Trans. Consum. Electron.}, 
  title={{UAV}-Enabled Fluid Antenna Systems for Multi-Target Wireless Sensing Over {LAWCNs}}, 
  year={Feb. 2026},
  volume={72},
  number={1},
  pages={1146-1155},
}

@ARTICLE{11281876,
  author={Yu, Kan and Li, Kaixuan and Zhao, Yujia and Feng, Zhiyong and Li, Dong and Zhang, Qixun and Yu, Jiguo},
  journal={IEEE Wireless Commun.}, 
  title={Movable Antenna-Aided Secure {V2X} Communication: An Integrated Sensing and Communication Perspective}, 
  year={Dec. 2025},
  volume={32},
  number={6},
  pages={118-124},
}

@ARTICLE{10354003,
  author={Zhu, Lipeng and Ma, Wenyan and Ning, Boyu and Zhang, Rui},
  journal={IEEE Trans. Wireless Commun.}, 
  title={Movable-Antenna Enhanced Multiuser Communication via Antenna Position Optimization}, 
  year={Jul. 2024},
  volume={23},
  number={7},
  pages={7214-7229},
}

@ARTICLE{10497534,
  author={Xiao, Zhenyu and Cao, Songqi and Zhu, Lipeng and Liu, Yanming and Ning, Boyu and Xia, Xiang-Gen and Zhang, Rui},
  journal={IEEE Trans. Wireless Commun.}, 
  title={Channel Estimation for Movable Antenna Communication Systems: A Framework Based on Compressed Sensing}, 
  year={Sep. 2024},
  volume={23},
  number={9},
  pages={11814-11830},
}

@ARTICLE{10654366,
  author={Liu, Wenchao and Zhang, Xuhui and Xing, Huijun and Ren, Jinke and Shen, Yanyan and Cui, Shuguang},
  journal={IEEE Wireless Commun. Lett.}, 
  title={{UAV}-Enabled Wireless Networks With Movable-Antenna Array: Flexible Beamforming and Trajectory Design}, 
  year={Mar. 2025},
  volume={14},
  number={3},
  pages={566-570},
}

@ARTICLE{11156108,
  author={Feng, Zhiyong and Zhao, Yujia and Yu, Kan and Li, Dong},
  journal={IEEE Trans. Commun.}, 
  title={Movable Antenna Empowered {PLS} With {Eve}’s Location Uncertainty: Joint Optimization of Beamforming and Antenna Positions}, 
  year={Dec. 2025},
  volume={73},
  number={12},
  pages={13708-13724},
}

@ARTICLE{11240557,
  author={Zhang, Xuhui and Liu, Wenchao and Ren, Jinke and Wang, Chunjie and Xing, Huijun and Shen, Yanyan and Cui, Shuguang},
  journal={IEEE Trans. Netw. Sci. Eng.}, 
  title={Movable-Antenna Empowered {AAV}-Enabled Data Collection Over Low-Altitude Wireless Networks}, 
  year={2026},
  volume={13},
  number={},
  pages={4506-4523},
}

@ARTICLE{11261377,
  author={Yang, Hanyu and Xing, Chengwen and Gong, Shiqi and Ju, Xin and Zhao, Nan and Niyato, Dusit},
  journal={IEEE Trans. Wireless Commun.}, 
  title={A Framework for Energy-Efficiency Optimization in {MA}-Aided {MU}-{MIMO} Systems}, 
  year={2026},
  volume={25},
  number={},
  pages={7551-7568},
}

@ARTICLE{11363226,
  author={Sun, Luyang and Wei, Zhiqing and Liu, Haotian and Yu, Kan and Li, Zhendong and Feng, Zhiyong},
  journal={IEEE Trans. Commun.}, 
  title={Movable Antenna-Assisted Flexible Beamforming for Integrated Sensing and Communication in Vehicular Networks}, 
  year={2026},
  volume={74},
  number={},
  pages={6149-6166},
}

@ARTICLE{10528324,
  author={Zuo, Yiping and Guo, Jiajia and Sheng, Biyun and Dai, Chen and Xiao, Fu and Jin, Shi},
  journal={IEEE Commun. Lett.}, 
  title={Fluid Antenna for Mobile Edge Computing}, 
  year={Jul. 2024},
  volume={28},
  number={7},
  pages={1728-1732},
}

@ARTICLE{11017619,
  author={Dou, Chenglong and Wu, Yuan and Qian, Liping and Wong, Kai-Kit and Quek, Tony Q. S.},
  journal={IEEE Wireless Commun. Lett.}, 
  title={Fluid Antenna Empowered Integration of Sensing, Communications and Computing With Hybrid Multi-Task Offloading}, 
  year={Aug. 2025},
  volume={14},
  number={8},
  pages={2521-2525},
}

@ARTICLE{11098592,
  author={Xiu, Yue and Zhao, Yang and Yang, Ran and Tang, Huimin and Qu, Long and Khabbaz, Maurice and Assi, Chadi and Wei, Ning},
  journal={IEEE Trans. Mobile Comput.}, 
  title={Latency Minimization for Movable Relay-Aided {D2D}-{MEC} Communication Systems}, 
  year={Jan. 2026},
  volume={25},
  number={1},
  pages={533-549},
}

@ARTICLE{11301741,
  author={Zhao, Jingwen and Chen, Ming and Yang, Zhaohui and Xu, Hao and Pan, Cunhua and Quek, Tony Q. S. and Wong, Kai-Kit},
  journal={IEEE Internet Things J.}, 
  title={Delay Efficient {FA}-Assisted Satellite Communication Network With Mobile Edge Computing}, 
  year={Mar. 2026},
  volume={13},
  number={5},
  pages={9436-9451},
}

@ARTICLE{11442647,
  author={Xiu, Yue and Zhao, Yang and Wang, Kaihe and Xu, Minrui and Niyato, Dusit and Liu, Guangyi and Wei, Ning},
  journal={IEEE Trans. Commun.}, 
  title={Delay Minimization for Movable Antennas-Enabled Anti-Jamming Communications With Mobile Edge Computing}, 
  year={2026},
  volume={74},
  number={},
  pages={6243-6258},
}

@ARTICLE{10620306,
  author={Chen, Pengcheng and Yang, Yuxuan and Lyu, Bin and Yang, Zhen and Jamalipour, Abbas},
  journal={IEEE Internet Things J.}, 
  title={Movable-Antenna-Enhanced Wireless-Powered Mobile-Edge Computing Systems}, 
  year={Nov. 2024},
  volume={11},
  number={21},
  pages={35505-35518},
}

@ARTICLE{10971386,
  author={Cang, Yihan and Chen, Ming and Yang, Zhaohui},
  journal={IEEE Commun. Lett.}, 
  title={Movable Antenna Enhanced {MEC} Systems With Discrete Antenna Position Selection}, 
  year={Aug. 2025},
  volume={29},
  number={8},
  pages={1749-1753},
}

@ARTICLE{11430678,
  author={Ju, Ying and Li, Mingdong and Wang, Haoyu and Liu, Lei and Qu, Youyang and Dong, Mianxiong and Leung, Victor C. M. and Yuen, Chau},
  journal={IEEE Trans. Mobile Comput.}, 
  title={Joint Channel Estimation and Computation Offloading in Fluid Antenna-assisted {MEC} Networks}, 
  year={Aug. 2026},
  volume={25},
  number={8},
  pages={12971-12989},
}

@ARTICLE{9076825,
  author={Bi, Suzhi and Huang, Liang and Zhang, Ying-Jun Angela},
  journal={IEEE Trans. Wireless Commun.}, 
  title={Joint Optimization of Service Caching Placement and Computation Offloading in Mobile Edge Computing Systems}, 
  year={Jul. 2020},
  volume={19},
  number={7},
  pages={4947-4963},
}

@ARTICLE{9419755,
  author={Zhao, Gongming and Xu, Hongli and Zhao, Yangming and Qiao, Chunming and Huang, Liusheng},
  journal={IEEE Trans. Parallel Distrib. Syst.}, 
  title={Offloading Tasks With Dependency and Service Caching in Mobile Edge Computing}, 
  year={Nov. 2021},
  volume={32},
  number={11},
  pages={2777-2792},
}

@ARTICLE{9861697,
  author={Xiao, Zhu and Shu, Jinmei and Jiang, Hongbo and Lui, John C. S. and Min, Geyong and Liu, Jiangchuan and Dustdar, Schahram},
  journal={IEEE Trans. Mobile Comput.}, 
  title={Multi-Objective Parallel Task Offloading and Content Caching in {D2D}-Aided {MEC} Networks}, 
  year={Nov. 2023},
  volume={22},
  number={11},
  pages={6599-6615},
}

@ARTICLE{10103637,
  author={Chu, Weibo and Jia, Xinming and Yu, Zhiwen and Lui, John C.S. and Lin, Yi},
  journal={IEEE Trans. Mobile Comput.}, 
  title={Joint Service Caching, Resource Allocation and Task Offloading for {MEC}-Based Networks: A Multi-Layer Optimization Approach}, 
  year={Apr. 2024},
  volume={23},
  number={4},
  pages={2958-2975},
}

@ARTICLE{9916163,
  author={Lyu, Zhonghao and Zhu, Guangxu and Xu, Jie},
  journal={IEEE Trans. Wireless Commun.}, 
  title={Joint Maneuver and Beamforming Design for {UAV}-Enabled Integrated Sensing and Communication}, 
  year={Apr. 2023},
  volume={22},
  number={4},
  pages={2424-2440},
}

@ARTICLE{11373590,
  	author={Jiao, Yingchao and Zhang, Xuhui and Liu, Wenchao and Wu, Yinyu and Ren, Jinke and Shen, Yanyan and Yang, Bo and Guan, Xinping},
  	journal={IEEE Trans. Veh. Technol.}, 
  	title={{UAV}-Enabled Data Collection for {IoT} Networks via Rainbow Learning}, 
  	year={Jul. 2026},
  	volume={75},
  	number={7},
  	pages={15433-15438},
}

@INPROCEEDINGS{10233456,
  author={Liu, Wenchao and Wang, Junyu and Xing, Huijun and Jin, Zhenzhen and Zhang, Xuhui and Shen, Yanyan},
  booktitle={Proc. IEEE/CIC Int. Conf. Commun. China}, 
  title={Blockchain-Empowered Space-Air-Ground Integrated Networks for Remote Internet of Things}, 
  year={Dalian, China, Aug. 2023},
  volume={},
  number={},
  pages={1-6},
}

@ARTICLE{11086503,
  author={Xue, Liang and Wang, Chunjie and Shen, Yanyan and Li, Shaobao and Wang, Shuqiang and Guan, Xinping},
  journal={IEEE Internet Things J.}, 
  title={{ETC}: Enhancing Transportation Computation With {IRS}-Enabled Wireless Powered {MEC} Systems}, 
  year={Oct. 2025},
  volume={12},
  number={19},
  pages={40837-40850},
}

@ARTICLE{10382559,
  author={Ma, Wenyan and Zhu, Lipeng and Zhang, Rui},
  journal={IEEE Commun. Lett.}, 
  title={Multi-Beam Forming With Movable-Antenna Array}, 
  year={Mar. 2024},
  volume={28},
  number={3},
  pages={697-701},
}

@ARTICLE{10592015,
  author={Zeng, Yaoping and Chen, Shisen and Li, Jinding and Cui, Yanpeng and Du, Jianbo},
  journal={IEEE Internet Things J.}, 
  title={Online Optimization in {UAV}-Enabled {MEC} System: Minimizing Long-Term Energy Consumption Under Adapting to Heterogeneous Demands}, 
  year={Oct. 2024},
  volume={11},
  number={19},
  pages={32143-32159},
}

@ARTICLE{10416363,
  author={Hu, Guojie and Wu, Qingqing and Xu, Kui and Si, Jiangbo and Al-Dhahir, Naofal},
  journal={IEEE Signal Process. Lett.}, 
  title={Secure Wireless Communication via Movable-Antenna Array}, 
  year={2024},
  volume={31},
  number={},
  pages={516-520},
}

@ARTICLE{11543384,
  author={Liu, Wenchao and Zhang, Xuhui and Ren, Jinke and Yuan, Weijie and You, Changsheng and Li, Shuangyang},
  journal={IEEE Trans. Commun.}, 
  title={{UAV}-Enabled {ISAC} With Fluid Antennas for Low-Altitude Wireless Networks}, 
  year={2026},
  volume={74},
  number={},
  pages={9529-9546},
}

@ARTICLE{11517458,
  author={Jiao, Yingchao and Zhang, Xuhui and Wang, Chunjie and Wang, Shuqiang and Shen, Yanyan and Ye, Kejiang and Xu, Chengzhong},
  journal={IEEE Trans. Consum. Electron.}, 
  title={{6DMA}-Enabled {ISAC} for Low-Altitude Economy}, 
  year={2026},
  volume={},
  number={},
  pages={},
}

@ARTICLE{11494046,
  author={Zhang, Xuhui and Wang, Chunjie and Liu, Wenchao and Xing, Huijun and Ren, Jinke and Xing, Zheng and Shen, Yanyan},
  journal={IEEE Trans. Consum. Electron.}, 
  title={Movable-Antenna Enabled Robust Vehicular Consumer Networks Under Imperfect {CSI}}, 
  year={Aug. 2026},
  volume={72},
  number={3},
  pages={6191-6201},
}

@article{xiao2025fundamental,
  title={Fundamental Models and Signal Processing for Movable Antenna-Enhanced Wireless Communications and Sensing},
  author={Xiao, Zhenyu and Pi, Xiangyu and Cao, Songqi and Zhu, Lipeng and Gao, Zhen and Xia, Xiang-Gen and Zhang, Rui},
  journal={Chinese J. Electron.},
  year={2026}
}

@ARTICLE{10741192,
  author={Xiao, Zhenyu and Pi, Xiangyu and Zhu, Lipeng and Xia, Xiang-Gen and Zhang, Rui},
  journal={IEEE Trans. Wireless Commun.}, 
  title={Multiuser Communications With Movable-Antenna Base Station: Joint Antenna Positioning, Receive Combining, and Power Control}, 
  year={Dec. 2024},
  volume={23},
  number={12},
  pages={19744-19759},
}

@ARTICLE{11226954,
  author={Zhu, Lipeng and Sun, He and Ma, Wenyan and Xiao, Zhenyu and Zhang, Rui},
  journal={IEEE Trans. Wireless Commun.}, 
  title={Multiuser Communications Aided by Cross-Linked Movable Antenna Array: Architecture and Optimization}, 
  year={2026},
  volume={25},
  number={},
  pages={6729-6744},
}

@ARTICLE{11418663,
  author={Zhu, Lipeng and Mao, Haobin and Ma, Wenyan and Xiao, Zhenyu and Zhang, Jun and Zhang, Rui},
  journal={IEEE J. Sel. Areas Commun.}, 
  title={Towed Movable Antenna ({ToMA}) Array for Ultra Secure Airborne Communications}, 
  year={2026},
  volume={44},
  number={},
  pages={3766-3781},
}

@ARTICLE{10896833,
  author={Chen, Ying and Yang, Yaozong and Wu, Yuan and Huang, Jiwei and Zhao, Lian},
  journal={IEEE Trans. Serv. Comput.}, 
  title={Joint Trajectory Optimization and Resource Allocation in {UAV}-{MEC} Systems: A Lyapunov-Assisted {DRL} Approach}, 
  year={March-Apr. 2025},
  volume={18},
  number={2},
  pages={854-867},
}

@ARTICLE{11197291,
  author={Xu, Wenhan and Pan, Cunhua and Yuan, Yulan and Wu, Yuan and Tsang, Danny H. K.},
  journal={IEEE Trans. Mobile Comput.}, 
  title={Reconfigurable Intelligent Surface Aided Mobile Fog Computing: A Space Aggregation-Based Lyapunov Driven Reinforcement Learning Approach}, 
  year={Mar. 2026},
  volume={25},
  number={3},
  pages={3809-3823},
}

@article{liu2025computing,
  title={Computing over Space: Status, Challenges, and Opportunities},
  author={Liu, Yaoqi and Han, Yinhe and Li, Hongxin and Gu, Shuhao and Qiu, Jibing and Li, Ting},
  journal={Engineering},
  volume={54},
  pages={20--25},
  year={2025},
  publisher={Elsevier}
}

@article{wang2023incentive,
  title={Incentive-aware blockchain-assisted intelligent edge caching and computation offloading for {IoT}},
  author={Wang, Qian and Chen, Siguang and Wu, Meng},
  journal={Engineering},
  volume={31},
  pages={127--138},
  year={2023},
  publisher={Elsevier}
}

@ARTICLE{10628028,
  author={Ren, Jinke and Zhang, Zezhong and Xu, Jie and Chen, Guanying and Sun, Yaping and Zhang, Ping and Cui, Shuguang},
  journal={IEEE Wireless Commun.}, 
  title={Knowledge Base Enabled Semantic Communication: A Generative Perspective}, 
  year={Aug. 2024},
  volume={31},
  number={4},
  pages={14-22},
}

\end{document}